\documentclass{article}
\usepackage{authblk}
\usepackage{natbib}
\usepackage[english]{babel}
\usepackage{orcidlink} 
\usepackage[letterpaper,top=2cm,bottom=2cm,left=3cm,right=3cm,marginparwidth=1.75cm]{geometry}
\usepackage{float}
\usepackage{pdfpages}
\usepackage{siunitx} 
\usepackage{amsfonts}
\usepackage{amssymb}
\usepackage{amsmath}
\usepackage{subcaption}
\usepackage{comment}
\usepackage{booktabs}
\usepackage{graphicx}

\def\drom{{\rm d}}

\def\rcore{R_{\rm core}}
\def\msuny{M_\odot~{\rm y}^{-1}}
\def\msuny9{10^{-9}~\msuny}

\title{The Crusts of Neutron Stars Revisited: Approximations within a Polytropic Equation of State Approach}

\author[1]{F. Köpp\,\orcidlink{0000-0001-9970-4339}}
\author[2]{J. E. Horvath\,\orcidlink{0000-0003-4089-3440}}
\author[3]{C. A. Z. Vasconcellos\,\orcidlink{0000-0002-9814-5317}}

\affil[1]{Departamento de F\'isica, CFM - Universidade Federal de Santa Catarina;
CEP 88040-900, Florianópolis, SC, Brazil. \it{fabiokopp@proton.me} }
\affil[2]{Universidade de S\~ao Paulo, Instituto de Astronomia, Geof\'isica e Ci\^encias Atmosf\'ericas,
	CEP 05508-900, S\~ao Paulo, SP, Brazil, \it{foton@astro.iag.usp.br}}
\affil[3]{Instituto de F\'isica, Universidade Federal do Rio Grande do Sul,
	Av. Bento Gon\c{c}̧alves 9500, 91501-970,
	Porto Alegre, RS, Brazil, and International Center for Relativistic
	Astrophysics Network (ICRANet),
	Pescara, Italy,
	\it{cesarzen@cesarzen.com}
}

\date{}
\begin{document}
\maketitle
\begin{abstract}
 In this work, we revisit several thin-crust approximations presented in the literature and compare them with the exact solutions of the Tolman–Oppenheimer–Volkoff (TOV) equations. In addition, we employ three different equations of state (EoSs) and one with pasta phase, each based on a distinct framework: the variational method, relativistic Brueckner–Hartree–Fock, and relativistic mean-field theory. We emphasize that these approximations require only the TOV solutions for the core and the EoS properties at the core–crust interface, in our approach only the energy density. Finally, the relativistic approximation, as well as the Newtonian approximation with corrections, show good agreement with the exact solutions. This means that a simple  treatment of the crust would suffice for structural purposes, independently of the 
 possible uncertainties in the sub-nuclear equation of state which are not very large. The unified EOS SINPA (relativistic mean-field theory) including the pasta phase was used to study the thin-crust approximation, while degeneracy in the \(M\)-\(R\) relation is demonstrated through: (i) anisotropic pressure in the modified TOV equation, (ii) \(f(R,L_m,T)\) gravity model, and (iii) dark matter admixture. As demonstrated, modifications to the description of gravitation introduce degeneracies in the mass--radius relation that are challenging to disentangle or quantify precisely.
\end{abstract}

\section{Introduction}
In the era of multi-messenger astronomy, observations such as the detection of gravitational waves from 
the GW170817 event\cite{LIGOScientific:2018cki}---marking the first observed merger of two neutron stars with an electromagnetic counterpart~\cite{LIGOScientific:2017vwq}. Together with key measurements of masses and radii by the Neutron Star Interior Composition Explorer (NICER) of the compact stars PSR~J0030+0451~\cite{riley2019nicer} 
\((M = 1.34^{+0.15}_{-0.16}\,M_\odot,\ R = 12.71^{+1.14}_{-1.19}\,\mathrm{km})\) 
and PSR~J0740+6620~\cite{riley2021nicer} 
\((M = 2.072^{+0.067}_{-0.066}\,M_\odot,\ R = 12.39^{+1.30}_{-0.98}\,\mathrm{km})\), 
have provided quite stringent constraints on the equation of state of dense matter. In particular, the combined tidal-deformability and mass--radius inferences disfavor several candidate EoSs reported in the literature (see, e.g.,~\cite{LIGOScientific:2017vwq}). Notably, these two pulsars, which exhibit nearly identical radii despite their significantly different masses (the former within \(1\sigma\)), challenge the status of many existing supranuclear EoS parametrizations. Along these lines, the authors of~\cite{Biswas:2021pvm} performed a Bayesian model selection among 31 equations of state and found that the most preferred ones are AP3 and MPA1. These EoSs predict the radius and the dimensionless tidal deformability of a $1.4\,M_\odot$ neutron star to be $12.10 \,(12.50)\,\mathrm{km}$ and $393 \,(513)$, respectively. 

The question of which EoS would be adequate to describe neutron star radii for given masses is entangled with the subnuclear expressions employed to integrate stellar models all the way down to the surface. For several years, the overall properties of the matter below nuclear saturation were considered as settled. However, the new 
measurements bring back the issue of how to construct a description embracing the measured values in which the subnuclear EoS is also involved. Moreover, the 
quest for an extended/alternative theory of gravitation affects both the supranuclear and the subnuclear regimes, not because of the composition, but rather when gravitation interplay eventually produces full stellar models. A variety of 
approximate models for the crust have been developed and considered useful in 
this context, as we shall see below. The solid crust of a neutron star (NS) with mass $M > 1\,M_\odot$ contains only a few percent of the total stellar mass. Despite its small contribution, the crust is believed to play a fundamental role in a variety of astrophysical phenomena. For instance, pulsar glitches are commonly attributed to the sudden unpinning of superfluid vortices present inside the inner crust; thermonuclear X-ray bursts are triggered by unstable burning of accreted matter on the crustal surface; and giant gamma-ray flares observed in magnetars have been associated with large-scale crustal fractures induced by magnetic stresses. Furthermore, torsional oscillations of the crust have been proposed as an explanation for quasi-periodic oscillations in magnetar flare tails. The thermal properties of the crust also influence the long-term cooling of isolated NSs as well as the post-outburst relaxation of X-ray transients, where heat deposited in the crust during accretion episodes gradually diffuses outward. For a comprehensive review, see  Ref.~\cite{Chamel:2008}.

The main motivation of this paper is to investigate whether accuracy in the theoretical description of the neutron star radii can be achieved by a detailed model of the stellar crust and the underlying core. For this purpose, we study firts a set of thin crust approximations (semi-analytical solutions) \cite{Zdunik:2016vza} and compare them with the numerical (exact) solutions of the TOV equations\cite{tolman1939,oppenheimer_volkoff1939} for the aforementioned EoSs. In addition to the previous work \cite{Zdunik:2016vza}, we analyzed the approximations for a fixed energy density (between two well-measured masses: 1.4, 2.0 (SINPA\cite{PhysRevD.106.023031}) and 2.08 $M_{\odot}$). 
Our results indicate that an $\sim 500 m$ uncertainty is intrinsic to the results for any approximation made for the crust. Therefore, the accuracy of measurements would need to reach the $\leq 100 m$ precision to pindown the nature of matter at subnuclear densities, a goal which is probably not that far away in time, and that will boost neutron star fine modeling. However, other effects may be important for this problem. In Appendix B, we calculated the anisotropic effect on the mass–radius (M–R) relation and the tidal dimensionless deformability using a quasi-local equation of state, considering only positive values of the anisotropy parameter and adopting the MPA1 equation of state. Furthermore, in order to quantify the modifications induced in the neutron-star sequence by the influence of modified gravity, we computed the M–R relation and the radial profile of the energy density (for a fixed stellar mass of 1.4 $  M_{\odot}  $) within the framework of $  f(R,L_m,T)  $ gravity, following the approach of Ref.~\cite{Mota:2024kjb} with $  L_m = p  $ and employing the unified SINPA\cite{PhysRevD.106.023031}) equation of state. The influence of admixed dark matter on the surface radius and the core radius is also briefly discussed  .

In addition, the results can be used to address the claim that the radius is essentially determined by the subnuclear equation of state, showing that this statement is partially true at most, and highlighting the dependence on the core radius.

\section{Setting of the problem}

\subsection{The nuclear and subnuclear EoSs}  

Motivated by these works of \cite{riley2019nicer, Biswas:2021pvm}, we have considered a neutron star described by the MPA1 \cite{Muther1987PLB} (based on relativistic Brueckner--Hartree--Fock calculations), the MS1 \cite{Mueller1996NPA} (based on relativistic mean field theory) EoS; the AP4 incorporating the Argonne two-nucleon interaction (AV18) with relativistic boost corrections $(\delta v)$ to the two-nucleon interactions and the Urbana model of three-nucleon interaction ($UIX^{*}$).\cite{Akmal1998PRC} based on variational methods); and finally the unified SINPA EoS modeling the so-called  pasta phase \cite{PhysRevD.106.023031}). All these EoSs are composed of $npe\mu$ composition for the cold matter.  
For a review of the aforementioned models, see \cite{Heiselberg:2000dn}.
The MPA1, MS1 e AP4 EoSs were generated using the framework (piecewise-polytropic EoS) presented in \cite{Read:2008iy}. All they support a maximum neutron star mass exceeding 2.08 $M_\odot$. In addition, in this framework,  the low-density EoS is fixed for $\rho = 10^{3} \, \text{g/cm}^3$ to $\rho = 10^{14} \, \text{g/cm}^3$ using four piecewise polytropic segments \cite{Read:2008iy}, adjusted to the SLy EoS \cite{sly4}. 
For the unified SINPA EoS, we employ the EoS table available at the repository\footnote{\url{https://github.com/hcdas/Unified_pasta_eos}}. Here, "unified" indicates that the same microscopic framework is used to describe both the crust and the core. A distinctive feature of this SINPA EoS is the incorporation of recent experimental atomic mass data from the AME2020 evaluation \cite{Huang_2021} to model the outer crust, as well as the inner crust (including pasta phases) via the compressible liquid drop model (CLDM).

As is well-known,The thin crust approximations derived here are applicable solely to catalyzed matter, which is considered a reliable approximation for isolated neutron stars. For accreted s, the composition is different (see the reference \cite{Zdunik:2016vza}) and some modification is expected.

 In the parametrization introduced by \cite{Read:2008iy}, the fit is constructed as follows: the first (lowest-density) segment of the piecewise-polytropic relation $p(\rho)$ is extrapolated to lower densities until it intersects the chosen low-density EoS. Further, the authors of \cite{ferreira2020neutron} studied the effects of matching procedures on density core-crust transition using meta-modeling for different nuclear symmetric parameters ($K_{sat}, Q_{sat}, E_{sym}, L_{sym}, K_{sym}$) for the core EoS while for the crusts, they used the Sly4 EoS. That worked compared two methods of matching procedure: $P(\epsilon)$ (the same as \cite{Read:2008iy}) and $P(\mu)$ (Maxwell's construction). The first method, denoted here as $P(\epsilon)$, simply enforces $P_{\text{crust}}(\epsilon) = P_{\text{core}}(\epsilon)$ at the core--crust interface, disregarding thermodynamic considerations. The Maxwell construction here denoted as $P(\mu)$ enforces both mechanical equilibrium and baryon-number conservation by imposing the following local condition at the phase transition $ P_{\text{crust}}(\mu) = P_{\text{core}}(\mu)$ where $\mu$ is the common baryon chemical potential. They found strong correlation between $L_{sym}$ and $K_{sym}$ for both matching procedures. Also, the mean value for $n_{cc}$ for $P(\mu)$ and $P(\epsilon)$ are $\sim 0.05$ and $\sim 0.072$ $fm^{-3}$, respectively. One can easily notice that core's size and crust length are mainly determined by matching procedures and ($L_{sym}$, $K_{sym}$) correlation as shown in table 2\cite{ferreira2020neutron}. The same work of the authors \cite{ferreira2020neutron} demonstrated that the crust-core matching method does not have a significant impact on $\Lambda$. 
In addition to that study of the matching procedures, a different analysis performed by the authors of \cite{PhysRevC.94.035804} shows that $P(\epsilon)$ is thermodynamically inconsistent. 
 
Besides the $P(\mu)$ and $P(\epsilon)$ methods, other thermodynamic methods are available \cite{GonzalezBoquera2017,Xu2009,GonzalezBoquera2019}. 
The latter methods involve identifying the region of mechanical and chemical equilibrium in stellar matter. The authors of \cite{Dutra2021} applied this method to relativistic mean-field models to investigate the effects of bulk parameters on the core-crust density transition. In addition, it is known that the location of the density transition is inversely proportional to the slope of the symmetry energy $L$ \cite{Horowitz:2000xj}. Using characteristic values of the nuclear symmetry energy slope parameter, $30 \leq L \leq 120$ MeV, the authors of \cite{Ducoin2011} employed two nuclear many-body approaches, namely Skyrme models and relativistic mean field calculations, to determine the core-crust transition density, finding $0.06 \leq n_{cc} \leq 0.10$ fm$^{-3}$.  The symmetry energy and its slope are constrained by both terrestrial and astrophysical experiments\cite{sym16081038}. Finally, some authors have employed an effective equation of state to describe the inner crust as $P(\epsilon) = A + B\epsilon^{\gamma}$\cite{Xu2009}, which will be addressed below.

It is important to emphasize that, in the subsequent sections, we have employed fixed nuclear parameters for the EoSs fitted to the piecewise polytropic approximation, whereas for the unified SINPA EoS we have adopted the tabulated EoS and constructed an analytical representation.

\subsection{A ladder of Crust Approximations}

In this subsection, we discuss the so-called thin crust approximations to the crust. These approximations are useful due to their simplicity and to semi-analytical expressions, as they require just interpolation methods to obtain the mass-radius diagrams. To use these approximations, it is necessary to solve the TOV equation for the EoS core and the complete EoS (crust + core), and determine the pressure (or density) at the crust-core transition. 

The Tolman-Oppenheimer-Volkoff (TOV) equation of hydrostatic equilibrium in General Relativity is\cite{Glendenning:1997wn}

\begin{equation}
\frac{{\rm d}P}{{\rm d}r}=-\frac{\rho G m(r)}{r^{2}}\left(\ 1+\frac{P(r)}{\rho(r) c^2}\right)\left(1-\frac{2Gm(r)}{rc^2}\right)^{-1}\left(1+\frac{4\pi r^{3}P(r)}{m(r)c^2}\right),
\label{eqn:TOV}
\end{equation}

where $m=m(r)$ is the gravitational mass enclosed within a sphere of radius $r$, $P$ is the pressure, and $\rho$ is the mass-rest density. The main point of the thin crust approximation is to assume that the mass of the crust, $M_{\rm crust}$, is small compared to the total mass $M$ of the neutron star. Thus, we assume $M_{\rm total} \sim M_{\rm core}$. In the following calculations we adopt  natural units, in which  $G = c = 1$.

For our purposes, we followed the considerations in Ref. \citet{Zdunik:2016vza}, neglecting the term $4\pi r^3 P/mc^{2}$ since it happens to be three orders of magnitude smaller than $\frac{P}{\rho c^2}$ at the neutron drip point. However, unlike the work of \citet{Zdunik:2016vza}, we assumed $\frac{P}{\rho c^2} \ll 1$.

In the next subsection, we presented four hierarchical crust approximations, namely, the simplest crust approximation, the Newtonian thin approximation,  the Relativistic thin crust (Tolman VII) and the Relativistic thin crust  approximation, all them well-known in the literature, and discuss later of their relevance for a proper evaluation of the neutron star radii. We further compare the aforementioned set of thin crust approximations with the exact TOV's solutions. As stated, the advantage of this approach is the semi-analytical solution which, when solved the TOV's for complete (for comparison with the semi-analytical result) and for the core EoS, It is possible to fully describe the solution of the TOV equations simply 
by interpolating the core mass to obtain the core radius and, thus, 
the stellar radius. The goal is to understand the role of 
each component and assess how uncertain the radii would be within them.
\newline

\subsection{Simplest crust app}

In this approximation the crust is just considered a slab atop the core, and the hydrostatic equilibrium equation is integrated in the Newtonian approximation (see below) . A correction coefficient simulating the effects of General Relativity ($\xi = 0.65$) is introduced by hand \cite{Horvath:1999gw} and the crust thickness becomes 

\begin{equation}
l_{\mathrm{simplest}} = \xi \frac{  \,  \, R_{\mathrm{core}}^{2}}{  \, \, M_{\mathrm{core}}}
\frac{\gamma}{\gamma-1}\frac{K\epsilon_{cc}^{\gamma }}{\epsilon_{cc}}
\end{equation}

where $K$ is the coefficient of the polytropic approximation of the adopted crust equation of state with index $\gamma$. To fully capture the density dependence in the inner crust, i.e., the stiffness (\(\gamma\)) for each energy density interval, a three-piecewise polytropic model is more suitable, as implemented in Ref. ~\cite{Read:2008iy}. Nevertheless, we show below that a single polytropic EoS fitted to the inner-crust EoS performs much better than expected when compared to the exact solutions.

It is important to emphasize that both $R_{\mathrm{core}}$ and $M_{\mathrm{core}}$ are obtained from TOV solutions for the core's EoS. In fact, this simple crust thickness is actually a limit from the derivation of the next subsection.

\subsection{Newtonian thin crust approximation derivation}

From equation (\ref{eqn:TOV}) with the assumptions aforementioned and neglecting the term $\left(1-\frac{2m}{r}\right)^{-1}$, we obtained,
\begin{equation}
\int^{\epsilon_{cc}}_{\epsilon_{surf}}\frac{ \frac{{\rm d}P}{{\rm d}\epsilon} {\rm d}\epsilon}{ \epsilon}=
-M_{\rm core} \int^{R_{\rm core}}_{R^{*}} \frac{\drom r }{ r^2 }~.
\label{newton}
\end{equation}

The left-hand side of equation (\ref{newton}) represents the contribution of the crust\footnote{Instead of expressing the equation in terms of a polytropic equation of state (EoS), it can also be written in terms of the chemical potential. For \( T = 0 \) (cold matter), the chemical potential is given by \( \mu = \frac{P + \epsilon}{n_b} \). Thus, 
$
\frac{dP}{\epsilon + P} = \frac{dP}{d\mu} \frac{d\mu}{\epsilon + P} = \frac{d\mu}{\mu}.
$
Integrating this equation yields \( \ln\left( \frac{\mu_{cc}}{\mu_0} \right) \), where \( \mu_{cc} \) is the baryon chemical potential at the crust-core interface, and \( \mu_0 = \mu(P=0) = m_0 \) is the energy per baryon at the neutron star  surface. The minimum energy is obtained for iron \(^{56}\text{Fe}\), with \( \mu_0 = 930.4 \, \text{MeV}\)\cite{Haensel:2007yy}. In our approach, since we are working in terms of a polytropic equation of state , we do not have an equivalent to $\mu_0$.}  for cold catalyzed matter (ground-state), and is given by 

\begin{equation}
    \int^{\epsilon_{cc}}_{\epsilon_{surf}}\frac{ \frac{{\rm d}P}{{\rm d}\epsilon} {\rm d}\epsilon}{ \epsilon} = \int^{\epsilon_{cc}}_{\epsilon_{surf}}\frac{ \frac{{\rm d} (K\epsilon^{\gamma})}{{\rm d}\epsilon} {\rm d}\epsilon}{ \epsilon}= \int^{\epsilon_{cc}}_{\epsilon_{surf}}  \gamma K\epsilon^{\gamma -2}   {\rm d}\epsilon = \frac{ \gamma}{\gamma - 1} (\frac{K\epsilon_\text{cc}^{\gamma}}{\epsilon_\text{cc}}  - \frac{K\epsilon_\text{surf}^{\gamma}}{\epsilon_\text{surf}} ) = \frac{ \gamma}{\gamma - 1} (\frac{P_\text{cc}}{\epsilon_\text{cc}}  - \frac{P_\text{surf}}{\epsilon_\text{surf}} ) 
\label{theta}
\end{equation}

The $\epsilon_{cc}$ refers to density energy at the core-crust interface and $\epsilon_{surf}$ to energy density at the star's surface. The last term of the equation (\ref{theta}) we denote as $\Theta$. The integrated right-side of the equation (\ref{newton}) is

\begin{equation}
    \frac{M_{\rm core}}{R_{\rm core}} - \frac{M_{\rm core}}{R^{*}}=\xi\Theta.
    \label{rint}
\end{equation}

Rearranging the surface radius ($R^{*}$) in the equation \ref{rint} , we have

\begin{equation}
 R^{*}_{\rm Newton} =  \frac{ R_{\text{core}}}{1 - \frac{\xi \Theta R_{\text{core}}}{M_{\text{core}}}}
\label{new_t}
\end{equation}
where $\xi$ is a correction factor used to approximate the relativistic solution.

The crust thickness $(l_{\rm crust}= R^{*}- R_{\rm core} )$ is now

\begin{equation}
    l_{\text{Newton}}= \frac{ R_{\text{core}}}{1 - \frac{\xi \Theta R_{\text{core}}}{M_{\text{core}}}} - R_{\text{core}} = \left( \frac{\xi R_{\text{core}}^2  \Theta}{M_{\text{core}}(1 - \frac{\xi \Theta R_{\text{core}}}{ M_{\text{core}}})} \right)  
    \label{l_n1}
\end{equation}
Expanding $(1-\frac{\xi \Theta R_{\text{core}}}{ M_{\text{core}}})^{-1}$ of equation~ (\ref{l_n1}), and assuming  $  1 \gg \frac{\xi \Theta R_{\text{core}}}{M_{\text{core}}}
$, we obtain \(  l_{\text{Newton}} = l_{\mathrm{simplest}} \). We tested the same factor $\xi = 0.65$ in equation~(\ref{l_n1}) without expansion, and found that the crust length and format are essentially the same as $l_{\mathrm{simplest}}$. However, our results use $\xi = 1$ for the crust length and $\xi = 0.55$ for the calculation of the $M - R$ relations.

In addition, the function \(\Theta\) is the same for all the approximations considered here. 
It becomes a constant when reproducing the \(M\)--\(R\) relation, since 
only the pressure and energy density at the core-crust interface 
(and \(P_{\text{surf}} = 0\)) are required. However, for the fixed energy density case ($r_{core}$ and  $m_{core}$ are fixed) and  $\Theta$  will change according to $(P_{cc}(\epsilon_{cc}) - P(\epsilon_{cc\rightarrow} \epsilon_{surf}))$, resulting in $l({\epsilon_{cc\rightarrow} \epsilon_{surf})}$ which is the crust length.

\subsection{Relativistic thin crust approximation}

In the relativistic thin crust approximation, the term $\left(1-\frac{2m}{r}\right)^{-1}$ is now considered. Then, we have,

\begin{equation}
\Theta = -M_{\rm core} \int_{R^{*}}^{R_{\rm core}} \frac{{\rm d}  r}{r^{2} \left(1 - \frac{2M_{\rm core}}{r}\right)}.
\label{rel}
\end{equation}
Solving the integral on the right side of equation (\ref{rel}) considering the following substitution $\mu=(1 - \frac{2M_{\rm core}}{r}$) , we obtain

\begin{equation}
 -\frac{1}{2} \int_{\mu(R^{*})}^{\mu(R_{\rm core})} \frac{d \mu}{\mu} =    -\frac{1}{2} \ln \left( \frac{1 - \frac{2M_{\rm core}}{R_{\rm core}}}{1 - \frac{2M_{\rm core}}{R^{*}}} \right)
 \label{der_rad}
\end{equation}
Then, equating (\ref{der_rad}) to $\Theta$, and rearranging for the stellar radius ($R^{*}$) yields  

\begin{equation}
       R^{*} =  \frac{2M_{\rm core}}{1 - \left(1-\frac{2M_{\rm core}}{R_{\rm core}}\right) e^{2\Theta(r)}},
       \label{rtotal}
\end{equation}

The crust thickness  for this approximation is then
\begin{equation}
    l_{\text{Relativistic}} = \frac{(1-e^{2\Theta(r)}) (2M_{\rm core}-R_{\rm core})}{1 - \left(1-\frac{2M_{\rm core}}{R_{\rm core}}\right) e^{2\Theta}},
\end{equation}
In this approximation, we evaluate two methods to determine \( R_{\text{core}} \). The first uses the Tolman VII solution, which results in
\[
R_{\text{core, thin}} = \left( \frac{3 \, M_{\text{core}}}{4 \pi \, \varepsilon_{cc} \, \left( \tfrac{2}{5} + \tfrac{3}{5} \, \frac{\epsilon_{0}}{\varepsilon_{cc}} \right)} \right)^{1/3},
\]
where \(\epsilon_{0}(n_{b} \simeq 0.16 \,\mathrm{fm}^{-3}) \simeq 145 \,\mathrm{MeV\,fm}^{-3}\) is the energy density at nuclear saturation density.  
The results of this approximation is denoted as \emph{Thin-Crust Approximation}, while the corresponding values of \( R_{\text{core}} \) obtained directly from the core TOV solutions are labeled \emph{Thin-Crust Approximation (Relativistic)}. We emphasize that both approaches are relativistic, differing only in the method used to determine \( R_{\text{core}} \).

It is also useful to obtain the fixed value of $\Theta$ using the following relation, 

\begin{equation}
     \Theta = \frac{1}{2} \ln \left( \frac{1-\frac{2M_{\rm core}}{R^{*}}}{1-\frac{2M_{\rm core}}{R_{\rm core}}} \right),
        \label{rc}
\end{equation}

\section{Crust Mass}
The crust mass can be estimated in a crude way integrating the following equation,

\begin{equation}
 \frac{dP}{dM}=- \frac{M}{4\pi R_{\rm core }^{4}(1-\frac{2M}{R_{\rm core }})  }
\end{equation}
thus,
\begin{equation}
 m_{\rm crust}  = \frac{4\pi R_{\rm core}^4 p_{t}}{ M_{\rm core}}\left(1-\frac{2M_{\rm core}}{R_{\rm core}}\right),
 \label{mcrust}
\end{equation}
where $p_{t}$ is the pressure at the crust-core interface. Equation~\eqref{mcrust} provides an excellent approximation, as demonstrated by the inset plot in Fig.~2 in Ref. \cite{Zdunik:2016vza}. The complete derivation of the $m_{\rm crust}$ can be found in Refs.  \cite{Zdunik:2016vza,PhysRevC.99.015803}.

\section{Stellar structure results}

To obtain quantitative results, the core equation of state was obtained from the complete EoS for density values greater than \( n_{cc} > 0.08~\text{fm}^{-3} \). As discussed below, this choice does not correspond to the actual 
transition density, which was determined within the framework used 
for the fit\cite{Read:2008iy}. The authors of \cite{Zdunik:2016vza} studied the dependence of the results on 
\( n_{cc} \). Values of \( n_{cc} \sim n_0 \) reduce the radius of the stellar sequence for masses below \( 1.4~M_{\odot} \) while the core's EoS solution also reduces its radius. Furthermore, it is well known that the slope of symmetry energy ($L$) affects \( n_{cc} \)  the crust length, the radius and the crust mass\cite{Dutra2021, GonzalezBoquera:2019yji,Lopes:2024bvz}. 

 Additionally, we compared a set of thin-crust approximations to exact numerical solutions of the Tolman-Oppenheimer-Volkoff equations for given fixed masses. The selected masses were \( 1.4~M_{\odot} \), \( 2.0~M_{\odot} \) and \( 2.08~M_{\odot} \). These solutions provided the crust-core energy density transition ($\epsilon_{cc}$) and the core radius ($R_{\rm core}$), both of which are required to evaluate the set of thin-crust approximations. In the profile results for fixed masses, the crust length is determined by the fixed core mass $  (M_{\rm core} ) $ and core radius $ ( R_{\rm core}) $, and depends on the energy density profile from $  \epsilon_{cc}  $ to $  \epsilon_{\rm surface}  $.

 We analyzed four equations of state: MPA1, AP4\footnote{The speed of sound in the medium for this AP4 EoS violates causality, i.e., $c_{\max}^{2} = 1.160$\cite{Read:2008iy}.}, MS1 and  the unified SINPA with pasta phase. For the reproduction of \( M - R \) relations, the values of \( \theta \) are comparable to \( H \)~\cite{LattimerPrakash2007} as well as the pressure transition.\footnote{For realistic EoSs, \( 1.04 \leq (H \text{ or } \theta) \leq 1.07 \) and \( 0.2~\text{MeV}/\text{fm}^{3} \leq p_{t} \leq 0.65~\text{MeV}/\text{fm}^{3} \).} Table~\ref{tab:core_crust} presents the properties of the core-crust transition for selected equations of state .

\begin{table}[H]
\centering
\caption{Properties of the core-crust transition for the selected EoSs.}
\begin{tabular}{lcccc}
\toprule
\textbf{EoS} & \textbf{$H$} & \textbf{$p_t$ (\si{MeV.fm^{-3}})} & \textbf{$\epsilon_t$ (\si{MeV.fm^{-3}})} & \textbf{$n_{cc}$ (\si{fm^{-3}})} \\
\midrule
MPA1 & 1.032 & 0.550 & 92.5 & 0.096 \\
AP4  & 1.036 & 1.040 & 115.0 & 0.122 \\
MS1  & 1.029 & 0.455 & 82.0 & 0.087 \\
SINPA & 1.029 (1.036 extracted) & 0.429 & 72.0 & 0.084   \\
\bottomrule
\end{tabular}
\label{tab:core_crust}
\end{table}

The EoSs MPA1 and MS1 were extensively analyzed in the context of the first direct detection of gravitational waves from the coalescence of a neutron star binary system, known as the GW170817 event\cite{LIGOScientific:2018cki}. Using credible intervals for the posteriors obtained from EOS-insensitive relations, a parametrized EOS without a maximum-mass requirement, and independent EoSs, it was shown in Fig.~1 of Ref. \cite{LIGOScientific:2018cki} that the MPA1 EoS lies within the 90$\%$ credible level of the marginalized posterior for the tidal deformabilities of the two binary components of GW170817. However, the MS1 EoS lies outside these credible intervals.

The value of $\epsilon_t = 50$~[MeV/fm$^3$] for the AP4 equation of state was multiplied by 2.3 to obtain the correct $H$ for the $M \times R$ relations. For the crust length, the transition energy density is $\epsilon_t = 50$~[MeV/fm$^3$], consistent with the transition pressure determined in Ref.\cite{LattimerPrakash2007}. The small difference between the values for the realistic EoSs can be attributed to the BBP EoS fit to a polytropic EoS, the framework used to obtain the EoS\cite{Read:2008iy} and the thin-crust approximation derivation for the polytropic EoS; differently from the Refs. \cite{Zdunik:2016vza,LattimerPrakash2007} for the MPA1, AP4 and MS1.
In the case of the unified SINPA EoS, we adopted the fit to the BBP equation of state (without a pasta phase) for the inner crust, while employing the core–crust transition energy density (or $n_{cc}$) taken directly from the tabulated SINPA EoS. A more consistent approach for comparing equations of state with and without a pasta phase can be found in Ref.~\cite{Providencia:2013dsa}, which employed the BPS EoS for the outer crust, Thomas-Fermi approximation to describe the inner , while the core EoSs was d by the non-linear Walecka models. In their Fig. 13, the surface radius - using the following fixed masses: 1.0, 1.44 and 1.6 $M_{\odot}$ - for the parametrization NL3, $NL3\omega\rho$, and FSU are basically the same with and without pasta. However, the parametrization IU-FSU still yields a small difference between the two cases.  

We should point out that the two EoS: DH and NL3$\omega\rho$ from the Ref. \cite{Zdunik:2016vza} have the following values for $H$ 1.024 and 1.026 respectively. In addition, the BBP fit slightly extrapolates the range covered by the tabulated EoS since $n_B > 0.08  ~ \text{fm}^{-3}$ which may introduce a small uncertainty in our results. One can indirectly conclude that even for a polytropic EoS fit to BBP EoS, the $\theta$ value almost matches the one from the EoS used here ( where the inner crust is the DH EoS~\cite{sly4}). 
In the following, we will present the results for the MPA1, AP4, MS1 and unified SINPA EoSs.
\vspace{1cm}
\subsection{The {\it{MPA1}} case}

\begin{figure}[H]
    \centering
    \begin{subfigure}{\textwidth}
        \centering
        \includegraphics[width=0.55\textwidth]{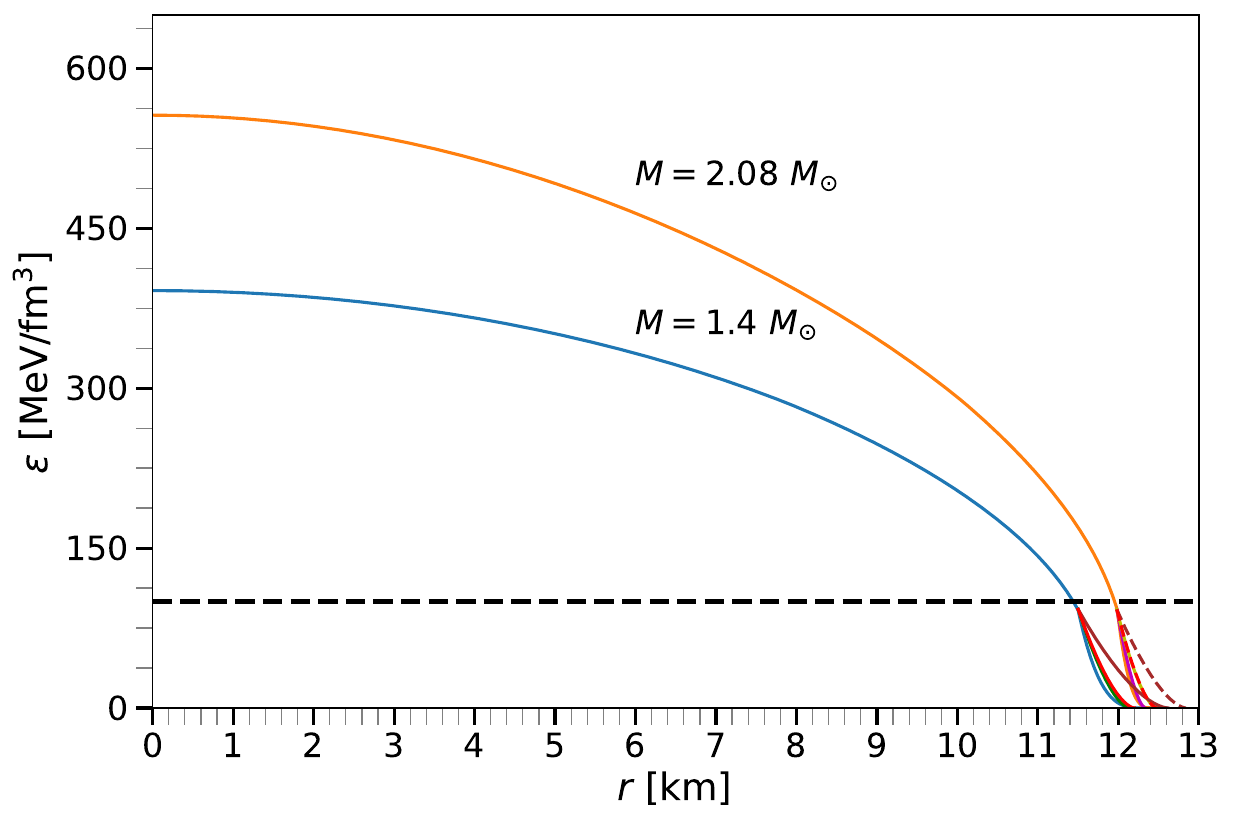}
        \caption*{Upper}
        \label{fig1}
    \end{subfigure}
    

 \begin{subfigure}{\textwidth}
        \centering
        \includegraphics[width=0.55\textwidth]{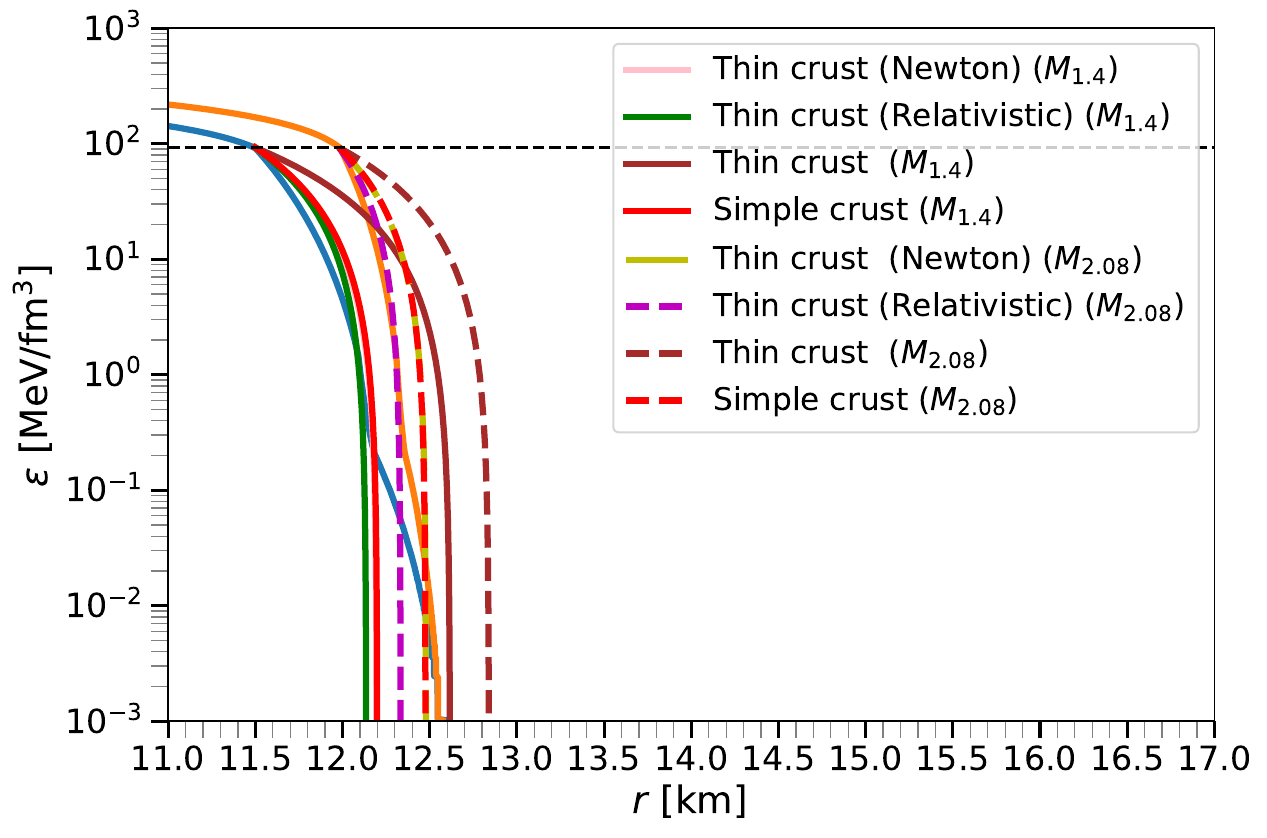}
        \caption*{Middle}
        \label{fig2}
    \end{subfigure}
  
\end{figure}

\begin{figure}[H]

    \begin{subfigure}{\textwidth}
        \centering
        \includegraphics[width=0.525\textwidth]{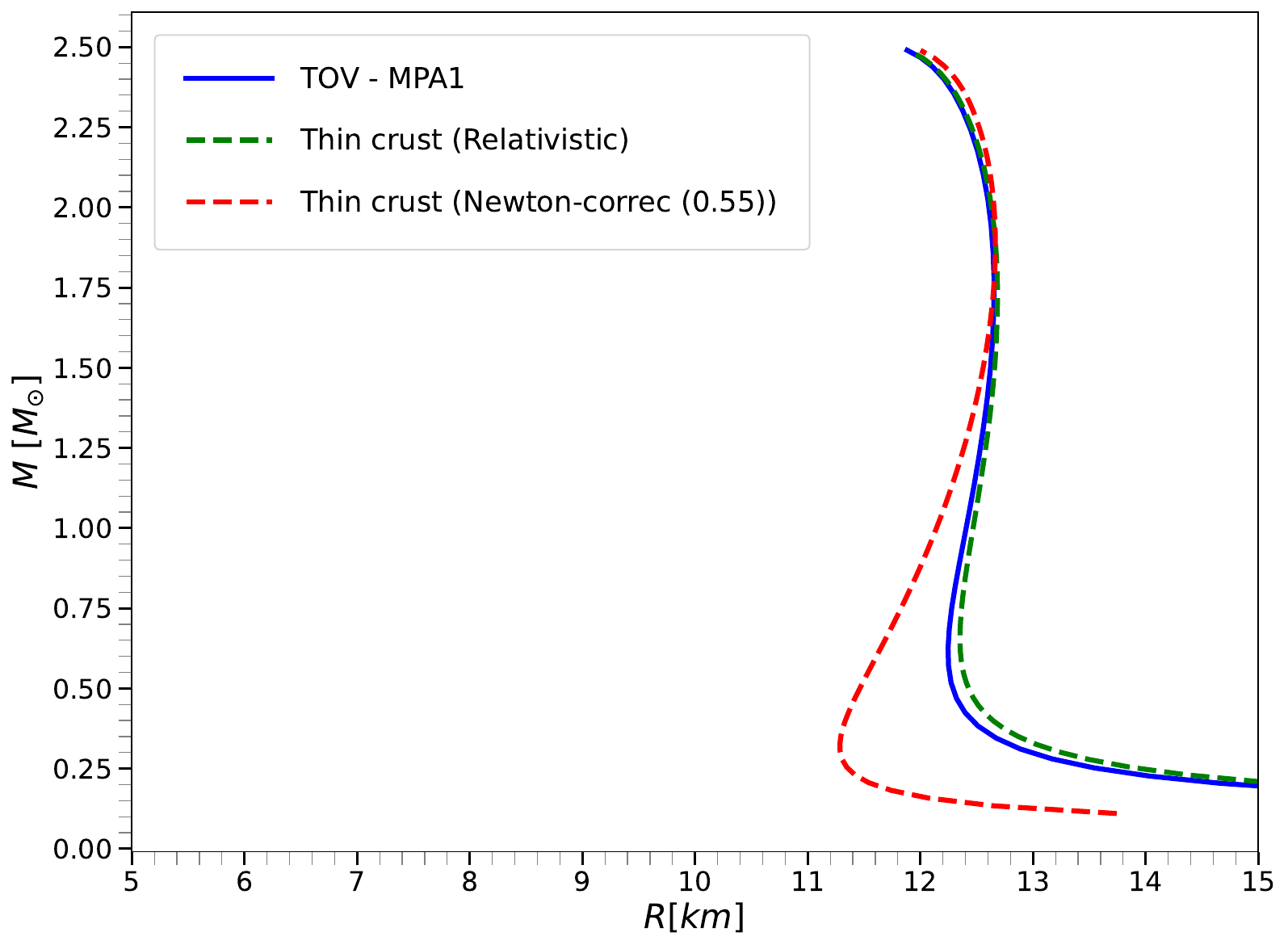}
        \caption*{Bottom}
        \label{fig3}
          \end{subfigure}

          \caption{Results for the EoS MPA1. Upper, density profiles for the $M=1.4 ~M_\odot$ and $M=2.08~M_\odot$ up to the surface. The dashed horizontal line marks the core-crust transition density. Middle, the simple approximation and the relativistic approximation is the most suitable for this EoS. As it stands, the simple approximation for $M=2.08~M_\odot $ has more deviation from the exact result than the one for $M=1.4 ~M_\odot$. Bottom, the blue solid line represents the exact TOV's solutions for the MPA1 EoS. The dashed green line, represents the thin crust approximation (relativistic). Even for an inner crust fit with just a polytropic EoS, the result is excellent. The red dashed line is the thin crust approximation for the Newtonian case with corrections. It is possible to see that even with correction, the results become accurate above $\sim \, M=1.4 ~M_\odot$. }
    \label{fig:mpa1_all}
    \end{figure}

\subsection{The {\it{AP4}} case}
\begin{figure}[H]
    \centering
    \begin{subfigure}{\textwidth}
        \centering
        \includegraphics[width=0.59\textwidth]{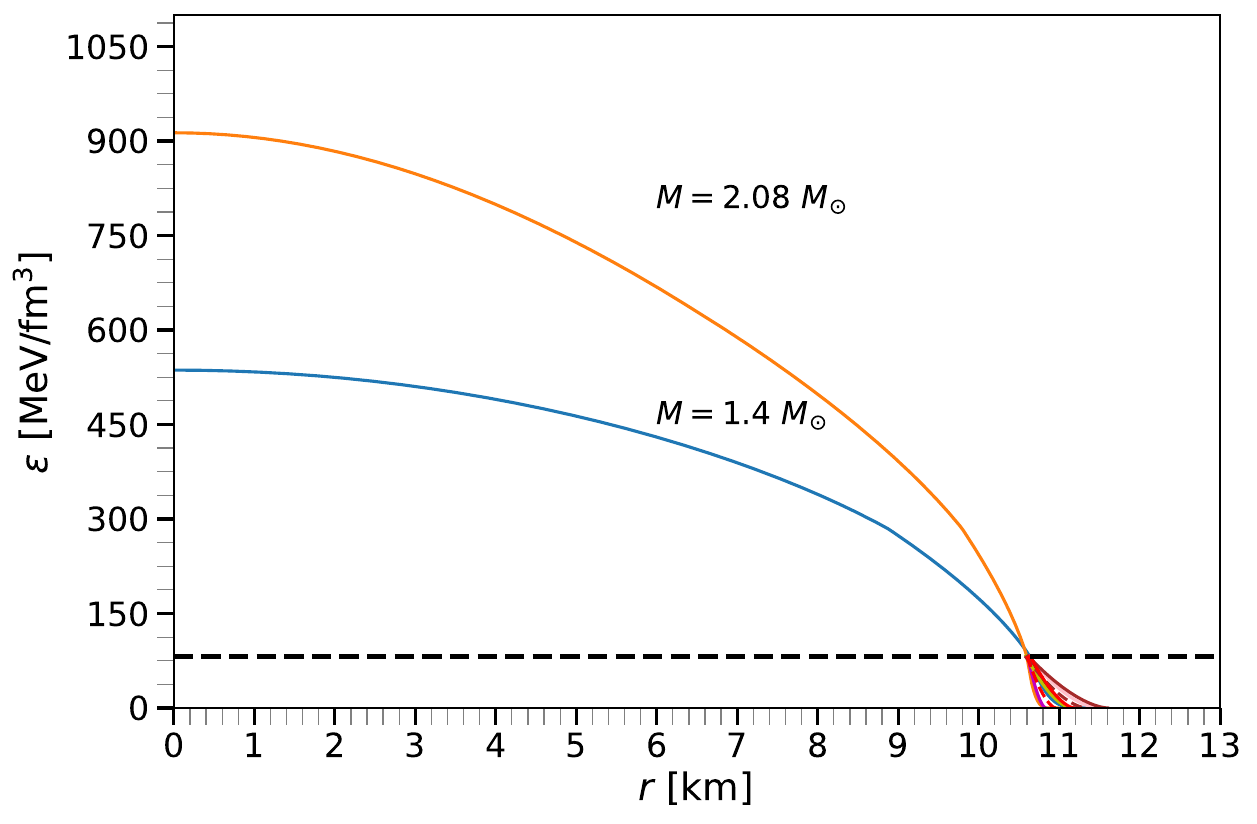}
        \caption*{Upper}
        \label{fig:epsilon-r-ap4}
    \end{subfigure}
    
    \begin{subfigure}{\textwidth}
        \centering
        \includegraphics[width=0.59\textwidth]{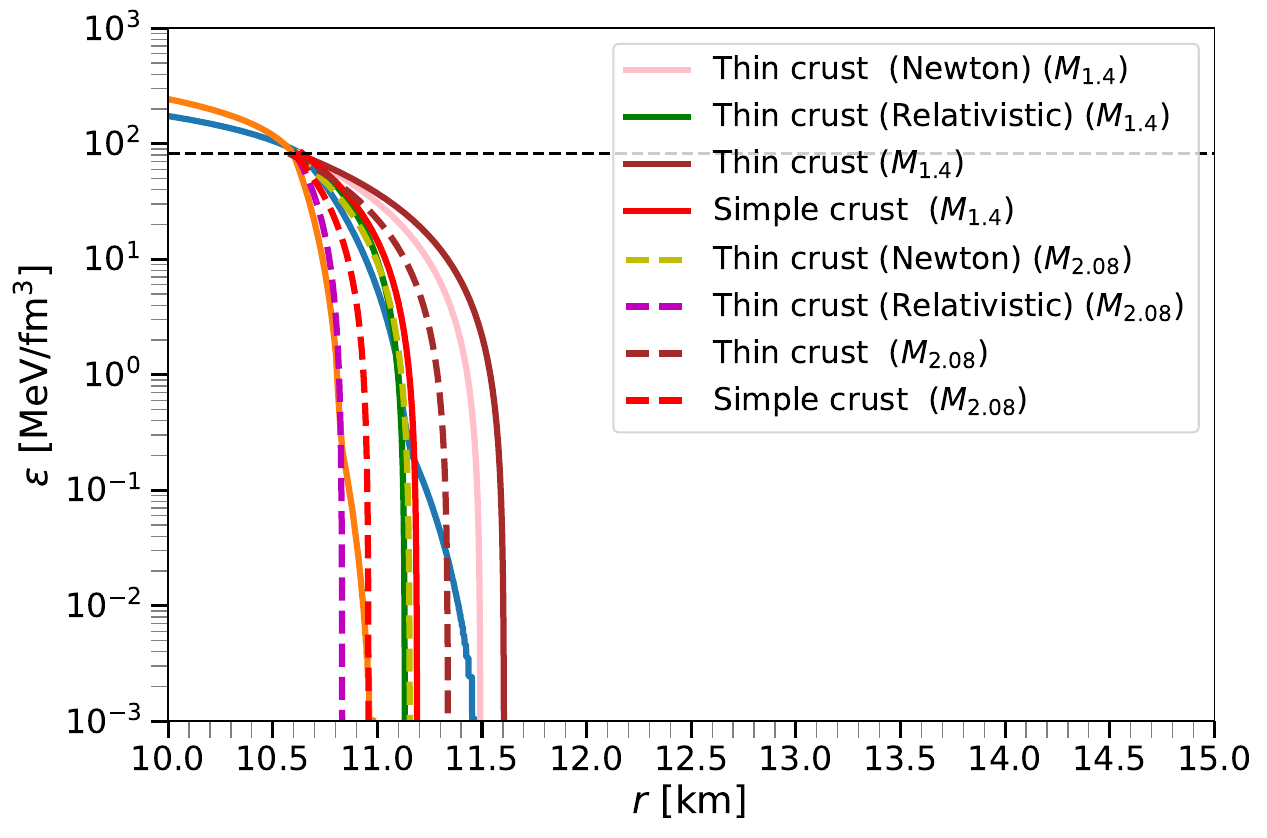}
        \caption*{Middle}
        \label{fig:epsilon-r-1-ap4}
    \end{subfigure}
    
    \begin{subfigure}{\textwidth}
        \centering
        \includegraphics[width=0.54\textwidth]{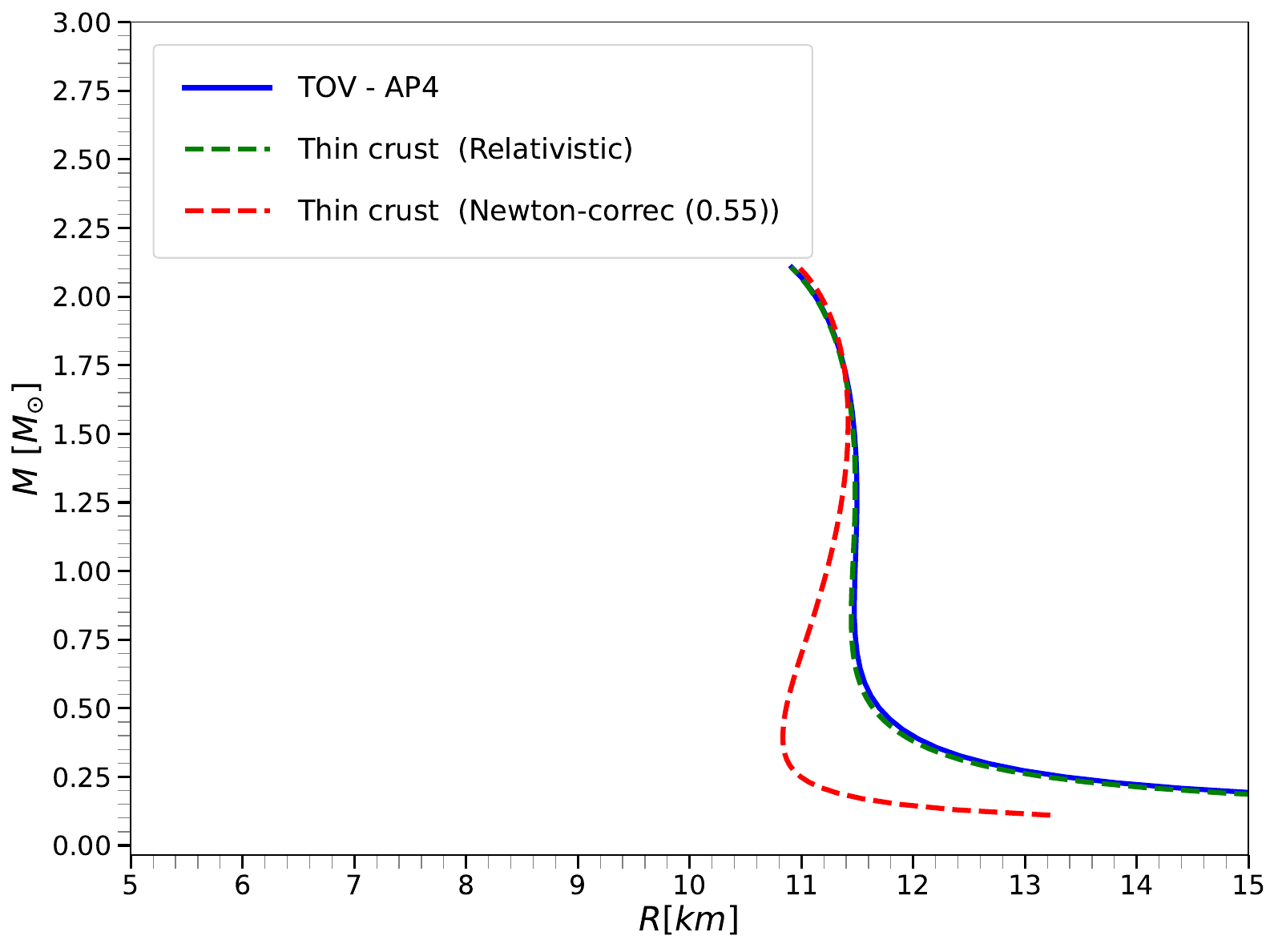}
        \caption*{Bottom}
        \label{fig:mr-sequence-tov-ap4}
    \end{subfigure}
    
    \caption{Results for the EoS AP4. Upper, the same as in  Fig.~\ref{fig:mpa1_all}. Note that the values for $\epsilon(r=0)$ for each NS mass are higher. Middle, the same as in Fig. \ref{fig:mpa1_all}, but now  the simple approximation for $M=1.4 ~M_\odot$ produces more difference in the radius than the same approximation for the $M=2.08 ~M_\odot$ model. Note that the radius of the cores are the same for this choice of the supranuclear equation of state. Bottom, the same as in Fig. \ref{fig:mpa1_all}.}
    \label{fig:ap4_all}
\end{figure}

\subsection{The {\it{MS1}} case}

\begin{figure}[H]
    \centering
    \begin{subfigure}{\textwidth}
        \centering
        \includegraphics[width=0.59\textwidth]{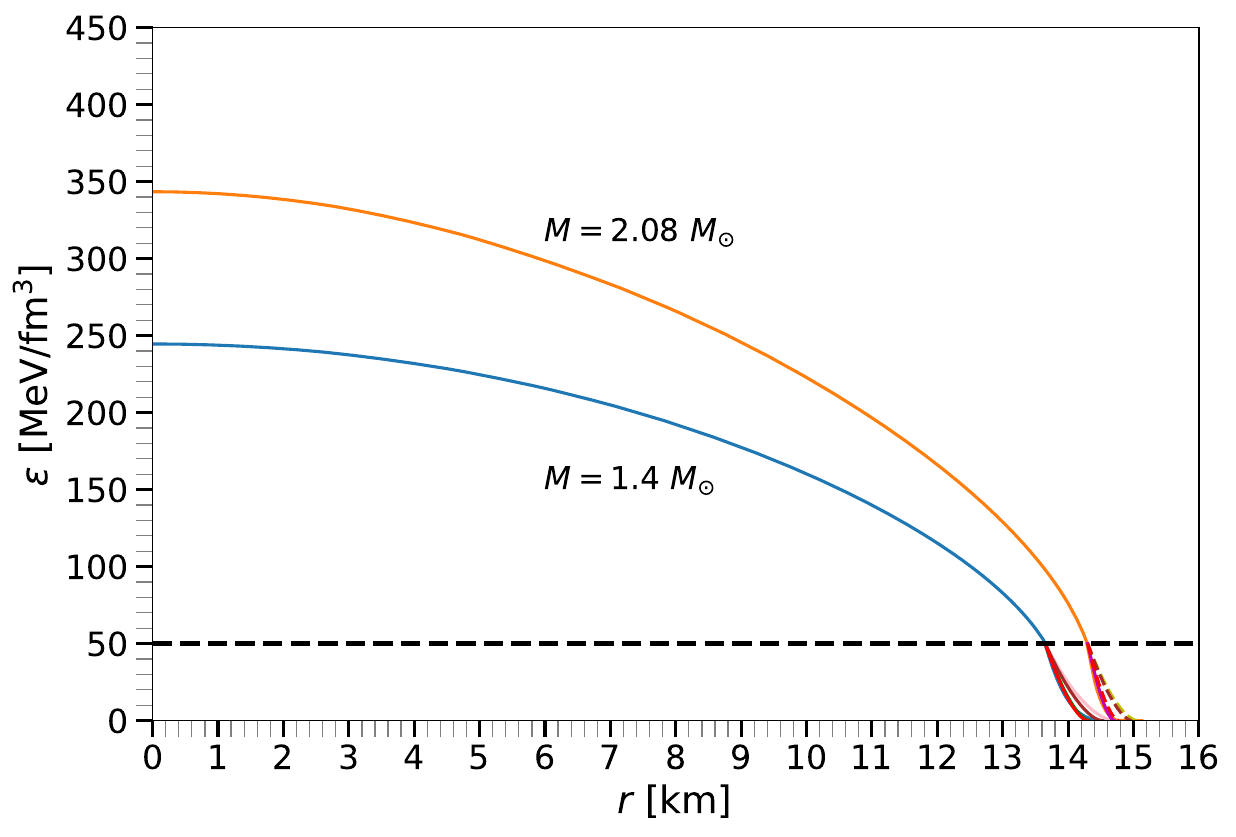}
        \caption*{Upper}
        \label{fig:epsilon-r-MS1}
    \end{subfigure}
    
    \begin{subfigure}{\textwidth}
        \centering
        \includegraphics[width=0.59\textwidth]{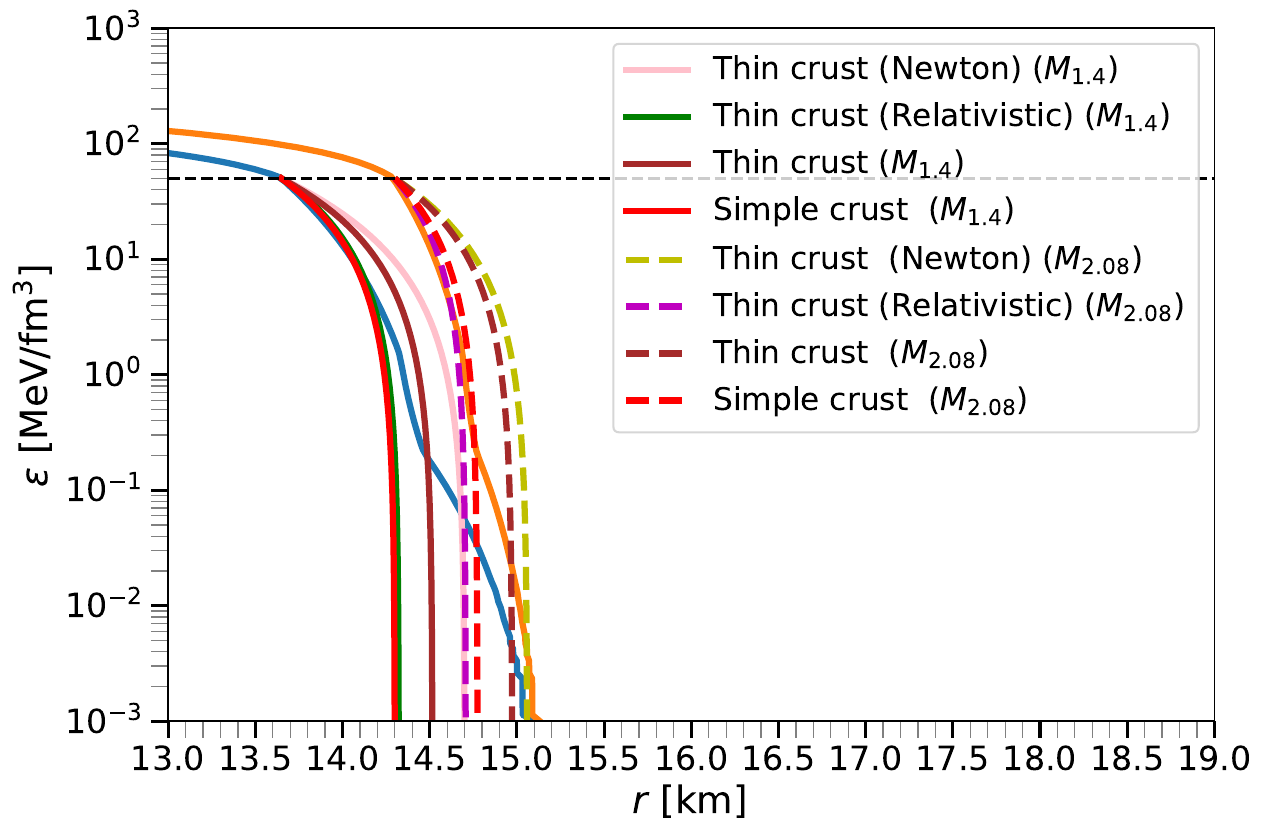}
        \caption*{Middle}
        \label{fig:epsilon-r-1-MS1}
    \end{subfigure}
    
    \begin{subfigure}{\textwidth}
        \centering
        \includegraphics[width=0.54\textwidth]{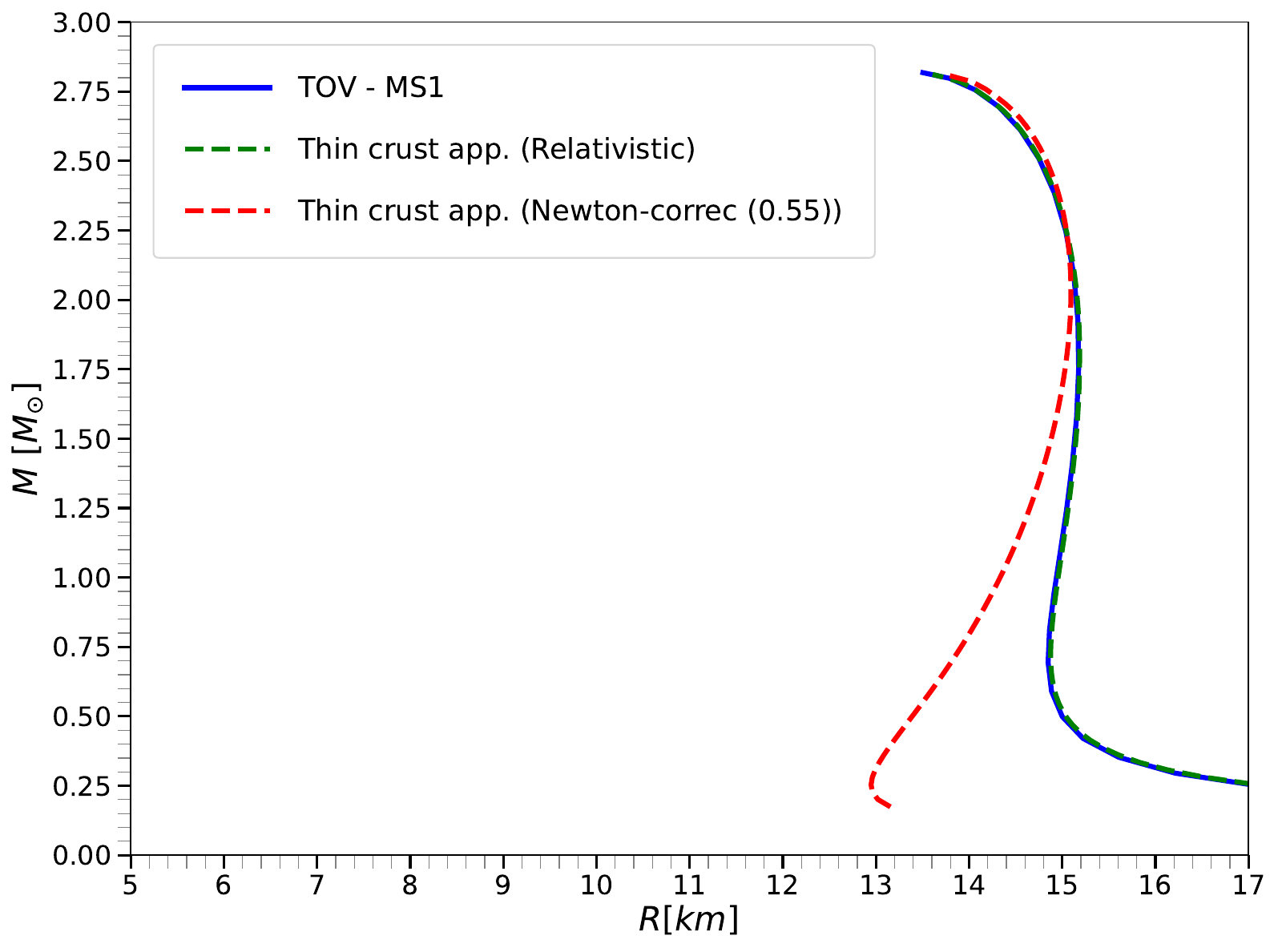}
        \caption*{Bottom}
        \label{fig:mr-sequence-tov-MS1}
    \end{subfigure}
    \caption{Comparison of results for the EoS MS1. Upper, the same as in Fig.~\ref{fig:mpa1_all}. Middle,  the same as in Fig. \ref{fig:mpa1_all}. Bottom, the same as in Fig. \ref{fig:mpa1_all}. The calculated radii are too large to provide a fair representation of the stars measured with the NICER data.}
    \label{fig:ms1_all}
\end{figure}

\subsection{The {\it{SINPA} } case}

As mentioned previously, the unified\footnote{The same EoS applied to both crust and core.} SINPA EoS exhibits a pasta phase and relies on recent AME2020 data\cite{Huang_2021} for the outer crust description. Our presented results assume no pasta in the inner crust (polytropic EoS fitted to BBP). Calculations were also carried out using the fitted SINPA inner crust. The resulting $M$-$R$ relations approach the no-pasta case below $\sim 1\,M_\odot$, which can be attributed to the slight increase of $H$ when fitting to the SINPA inner crust\footnote{It can be concluded that a polytropic fit is not really appropriate for this case.}. The energy density as a function of radius yields very similar results 
for both the total mass for the relativistic thin-crust case when using either of the two fits for the inner crust, namely BBP  and SINPA. This similarity arises because 
both fits - BBP and SINPA -\footnote{The BBP EoS is characterized by $K = (3.75785756 \times 10^{-4} \pm 1.46 \times 10^{-5}) \, 
\text{MeV} \, \text{fm}^{-3(1-\gamma)}$ 
and an adiabatic index 
$\gamma = 1.61131624 \pm 9.46 \times 10^{-3}$, 
while for SINPA we have 
$K = (4.83486099 \times 10^{-4} \pm 1.46 \times 10^{-5}) \, 
\text{MeV} \, \text{fm}^{-3(1-\gamma)}$ 
and $\gamma = 1.54633998 \pm 7.13 \times 10^{-3}$.}. yield similar pressures beyond $n_{cc}$.
The crust-core transition density in the case without pasta phases 
(BPS-BBP EoSs) is $n_{cc} = 0.08 \, \text{fm}^{-3}$. For the complete SINPA EoS 
the transition density becomes $n_{cc} = 0.0836 \, \text{fm}^{-3}$.

The transition density depends on the bulk, surface, curvature, and Coulomb energies\footnote{In the CLDM, the cluster energy is defined by the sum of these terms.}, as detailed in the section on the inner crust in Ref. \cite{Parmar:2021dbu}.   

\begin{figure}[H]
    \centering
    \begin{subfigure}{\textwidth}
        \centering
        \includegraphics[width=0.59\textwidth]{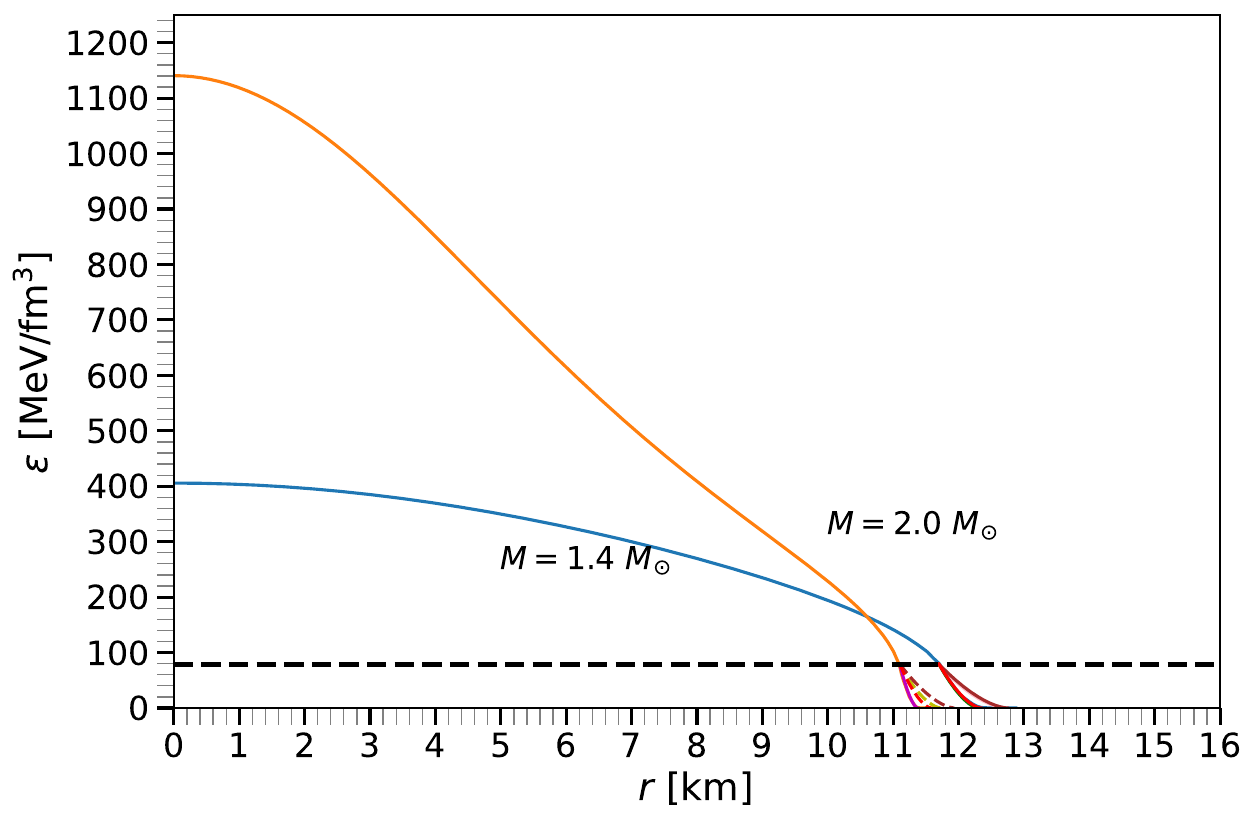}
        \caption*{Upper}
        \label{fig:epsilon-r-sinpa}
    \end{subfigure}
    
 \begin{subfigure}{\textwidth}

 \centering
        \includegraphics[width=0.59\textwidth]{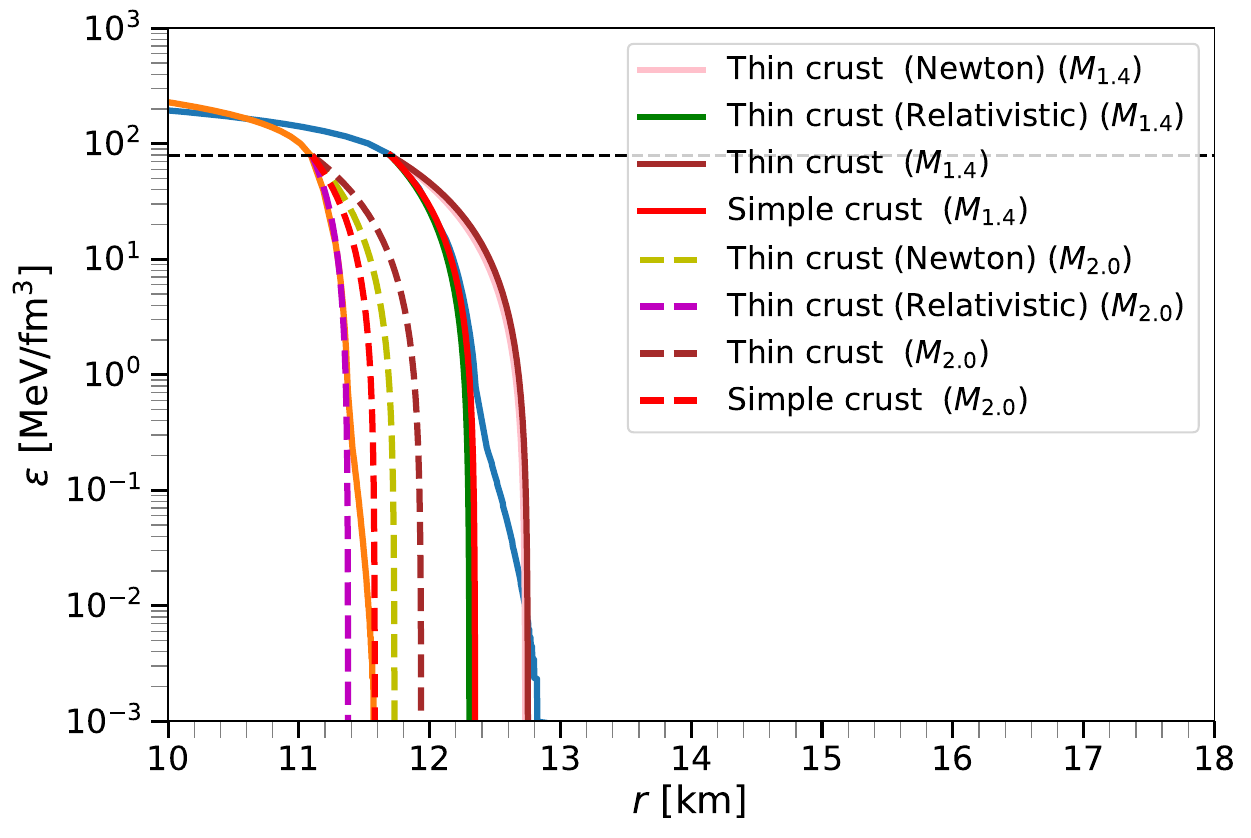}
        \caption*{Middle}
        \label{fig:epsilon-r-1-sinpa}
    \end{subfigure}
\end{figure}

\begin{figure}[H]

    \begin{subfigure}{\textwidth}
        \centering
        \includegraphics[width=0.54\textwidth]{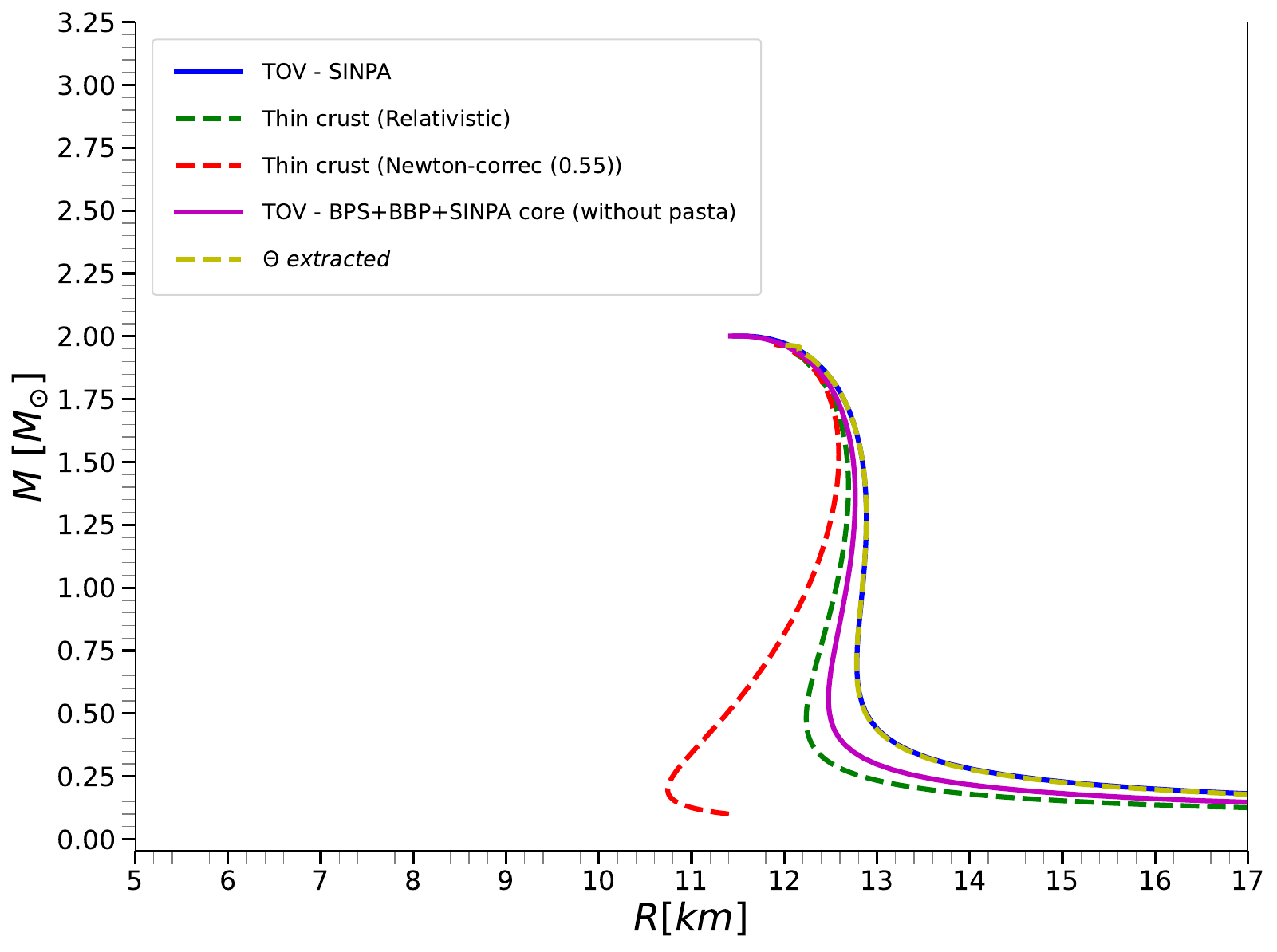}
        \caption*{Bottom}
        \label{fig:mr-sequence-tov-sinpa}
    \end{subfigure}
\caption{Results obtained with the SINPA EoS. 
Upper, same as Fig.~\ref{fig:mpa1_all}, but this EoS does not allow neutron stars with $M > 2.0\,M_{\odot}$  and the inner crust treated without a pasta phase (BBP fit). 
Middle, the same as Fig.~~\ref{fig:mpa1_all}. However, the radius difference between exact TOV - with pasta - solutions and relativistic thin-crust approximations  (BBP fit for inner crust with no pasta phase) remains below 400\,m. 
Bottom,  the same as Fig.~~\ref{fig:mpa1_all}. Mass-radius relation: exact TOV solution (solid blue), TOV without pasta (BPS+BBP+SINPA core; solid magenta), relativistic thin-crust approximation without pasta (BBP fit; green dashed), Newtonian thin-crust approximation without pasta (BBP fit; red dashed), and relativistic thin-crust approximation using $\Theta$  value from Equation~\eqref{rc} (yellow dashed). Since $\Theta$ is fixed from the SINPA EoS for both core and full model, the approximation closely matches the exact TOV result (solid blue line). 
  }
    \label{fig:combined-results}
    \end{figure}
\begin{figure}[H]
    \centering
    \includegraphics[scale=0.6]{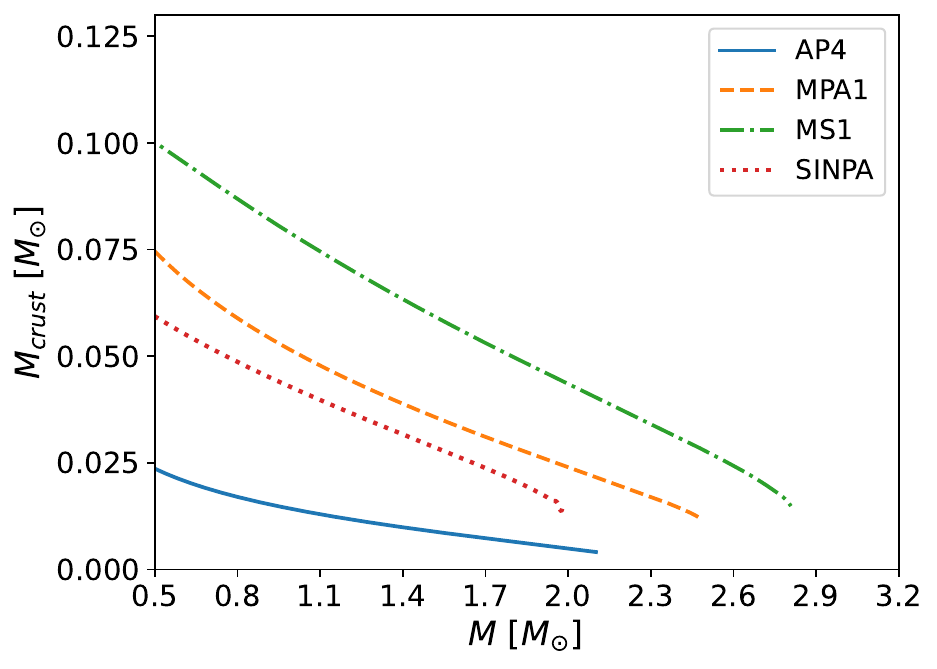}
    \caption{The mass of the crust as a function of neutron star mass for the four equations of state, calculated using Eq.~\eqref{mcrust}: MPA1 (blue), MS1 (orange), AP4 (green), and unified SINPA (red). Our results agree with those shown in Fig.~1(a) in the Ref.~\cite{Dutra2021} at comparable transition pressures~$p_t$. For the unified SINPA equation of state, the present approximation reproduces the crustal mass at~$2\,M_{\odot}$ reported in Table~III in Ref.~\cite{PhysRevD.106.023031}. }
    \label{fig:12}
\end{figure}

With these observations and calculations at hand, it is worth to reinforce the whole picture by looking at the crust mass. We depict in figure~\ref{fig:12} the whole range of masses for each of the equations of state. The most massive crust is just $\sim 0.08 M_{\odot}$ for 1.0 $M_{\odot}$, and in general much lower values are obtained. This justifies why the approximations work so well, which in turn is a warning about the real meaning of detailed calculations when confronted to real data, as stated above.

\section{Moving forward: glitches as decisive tools to study the crust}

More than 50 years have elapsed since the first {\it glitches} of the Cab and Vela pulsar were detected \cite{Anto}. These sudden spin-up of the pulsar frequency did not produce any appreciable change in the pulse, and therefore were associated to some internal factor. In addition to the prompt {\it starquake \, model} \cite{Ruderman, BaymPines}, a promising and exciting explanation was suggested to reflect the dynamics of (macroscopic) superfluids in the crust, a sudden decoupling and recoupling on a variety of timescales \cite{PinesAlpar}. However, a variety of events akin to glitches were identified over the years, which include slow glitches \cite{Shabanova}, anti-glitches \cite{Archibaldetal} and many others. In addition, the cracking of the crust and the motion of quantized vortices in the superfluid crust component have been elaborated and they are not separate explanations any more \cite{Anna}. The role of the core components have been generally neglected, although it is believed that a core superfluid could be involved at some level, at least in a subset of events \cite{Thapa, Anto}. We shall focus below on a few features that seem particularly interest and may shed light on the crust problem discussed above.

The superfluid component is a substantial fraction of the whole crust, and specific models can calculate how much. But this depends in turn on the specific form of the interactions at these densities, which are expected to yield energy gaps of the order of $\sim few \%$ of the neutron mass, i.e., not large to affect the structural properties.

Nevertheless, the knowledge of the moment of inertia in this superfluid component may be revealing provided the glitch model itself is correct \cite{Dutra2021}. One important discussion raised by Andersson et al. \cite{Nils} is related to the so-called {\it entrainment coupling} of the superfluid 
vortices with the nuclear lattice in the crust. The main point is that this effect may produce a reduced fraction of the neutron superfluid to decouple in a glitch, and the associated moment of inertia may not be enough the give a consistent picture of these events. While further work rejected this possible entrainment \cite{NoEntrain}, the determination of the mass, radii and moment of inertia of the star and the crust depend crucially on its presence or absence. In particular, the moment of inertia could be a probe of these interactions if a systematic analysis of theoretical EoS could be performed. For instance, Dutra et al. \cite{Dutra2021}
concluded that the moment of inertia is enough if relativistic mean field theory parametrizations are employed. A more recent work \cite{Japas} attempted to tie the nuclear forces to the phenomenology In summary, more detail studies could reveal important clues about the issues raised in our former sections. We believe that structural calculations alone will not allow to peep in the 
crust and core states alone, unless they can be combined and complemented with dynamical studies 
of the type made in Ref. \cite{Japas}, and this is the basic message of the present work which showed the type of degeneracy related to the crust, the gravitation theory underneath and other 
ingredients (see Appendix B).

\section{Conclusions}

The determination of the properties of matter at sub and supra nuclear densities is one of the key ingredients to 
build a consistent picture of neutron stars and their behavior. While it is often stated that the physics is well-known  below the nuclear saturation, considerable uncertainty still remains, and this is why a great deal of activity in the field 
is seen. For example, the presence of structures generically termed ``pasta'' \cite{russos, Deb,Haensel:2007yy,Providencia:2013dsa} near the core-crust 
transition described above within one EoS is still under debate. Accurate equations of state based on the best nuclear physics available have 
been constructed \cite{Constanca}, but even if we could trust them completely, an observational confirmation is 
not yet possible because the stellar structure can not be determined completely. This is one of our main points:
from the analysis of the results, and comparing the thin crust approximations with their exact numerical values, we see that it is not possible today to disentangle effects of an accurate 
equation of state from the treatment of the structure (i.e., approximate vs. exact). 
In Ref.~\cite{Carreau:2019zdy}, the authors perform a Bayesian analysis using a unified meta-modeling approach for the nuclear equation of state - without pasta - to investigate the correlations between low-density and high-density parameters and their impact on the crust-core transition, employing a compressible liquid-drop model. The surface tension emerges as the parameter exerting the strongest influence on the crust-core transition density. When the isospin dependence of the surface tension (parameter p) is constrained to a reasonable value, strong correlations are recovered between the transition properties an the isovector EoS parameters, specifically $  L_{\rm sym}  $, $  K_{\rm sym}  $, and $  Q_{\rm sym}  $. Bayesian analysis represents one of the most effective approaches for determining the crust-core transition density in neutron stars, as it systematically integrates theoretical predictions and experimental constraints from nuclear equations of state with observational astronomical constraints.
Other issues like addressing the existing alternatives for the 
theory of gravitation (Appendix B) to be applied in the neutron star structure would complicate even more these issues. Note that 
the simple approximations were performed assuming that General Relativity is correct, something which is under 
scrutiny and prompted the alternative gravitation theories currently in progress.

In addition to this point, we argue that the statement that the radius is essentially determined by the subnuclear equation of state and the mass by the supranuclear one is only partially true. Fig. 1, 4 and 7, and equation (10) 
show that the {\it core} radius introduces an uncertainty which is at least comparable to the one stemming from 
the approximate structure of the crust. Since both quantities enter the final value of the radius, we conclude that unless the uncertainty measurements of the radii can 
be reduced to $\sim 100 \, m$ \cite{BaoAn}, the detailed modeling of neutron stars will continue to suffer from that level of indeterminacy. In other words, we must enter a new era of precision measurements 
to confidently proceed with a detailed evaluation of the dense matter behavior. Again, if the theory of gravitation is questioned, and alternative models are introduced, that level of uncertainty would be reflect in a 
degeneracy of the models, unless the equation of state and the gravitation effects could be 
disentangled by some independent reliable procedure. In Appendix B, calculations are performed for both anisotropic pressure in the case of spherically symmetric neutron star and for the $  f(R,L_{m},T) = R + \alpha T L_{m}  $ gravity model\footnote{The analytical representation of the unified SINPA EoS employed in these calculations is presented in Appendix C.}. These calculations illustrate degeneracies between nuclear matter models and modified gravity theories, demonstrating how both approaches can still adequately describe the available astrophysical data.  Furthermore, the degeneracy effects induced by dark matter on neutron star properties are examined.

\section{Acknowledgements}{
This work has been done as a part of the Project INCT-Física Nuclear e Aplicações, Projeto No. 408419/2024-5. JEH wants to thank the CNPq Agency (Brazil) and the FAPESP Agency (S\~ao Paulo State) for continuous financial support over the years.
}

\appendix

\section{Baym-Bethe-Pethick (BBP) EoS FIT}

The interior of a neutron star, composed of cold, dense matter, consists of multiple shells, ranging from a solid crust to a liquid core. The crust is generally divided into an outer crust and an inner crust, separated by the neutron drip density. The outer crust comprises a lattice of fully ionized nuclei embedded in a degenerate electron gas, with its properties weakly dependent on density. For instance, in the seminal paper \cite{Baym1971BPS}, the equation of state (EoS) of the outer crust is obtained by minimizing the total energy of the system, $E_{\mathrm{tot}}$, per unit volume at fixed baryon number density $n_b$, for nuclei with mass number $A$ and atomic number $Z$, prior to neutron drip. This minimization is carried out using nuclear mass tables to identify, at each fixed $n_b$, the nucleus $(A,Z)$ that yields the lowest $E_{\mathrm{tot}}$.
  The adiabatic index ($\gamma$) is approximately $4/3$  as one can see in Table 5 in Ref. \cite{Baym1971BPS} for the mass density 
$10^{8} \, \text{g\,cm}^{-3} \lesssim \rho \lesssim 4 \times 10^{11} \, \text{g\,cm}^{-3}$, also in the same table, 
 the adiabatic index suffers a dramatic drop at neutron drip threshold. After this energy density, the adiabatic index increases in almost inner crust region. That behavior is influenced by several factors: the stiffening effect from interactions among dripped neutrons, the softening effect due to the coexistence of neutron gas and nuclear matter, and the softening contribution from Coulomb interactions. At the crust-core interface, the adiabatic index ($\gamma$) increases sharply from approximately 1.7 to 2.05, corresponding to the dissolution of nuclear structures. For an analytical representation of a unified equation of state based on the Brussels-Montreal models (BSk19, BSk20, and BSk21), as well as the Fortran routine fitted for them, see in Ref. \cite{Potekhin:2013qqa}. A comparison between the ``old'' outer crust and inner crust models with their recent versions, together with a detailed discussion of the unified EoS modelling, can be found in Ref. \cite{Sharma:2015bna}.  

Furthermore, the inner crust may have non-spherical structures such as cylindrical clusters (rods), slabs, cylindrical holes (tubes), and spherical holes (bubbles) may form, distinguishing these ``pasta phases'' from model dependencies remains a challenge. For example, Fig.~1 in \cite{refId0}, where the authors extended the CLDM to include pasta phases, illustrates the equilibrium of these phases and their corresponding transition densities between geometries for various nuclear functionals. Notably, both the crust-core transition and the baryon density occupied by each geometry are model-dependent (surface and curvature parameters as well as energy functionals).

To model the inner crust of neutron stars using a polytropic equation of state, defined as \(P(\epsilon) = K \epsilon^{\gamma}\), 
 the BBP\footnote{Nuclei are modeled within the framework of a compressible liquid-drop model (CLDM).} EoS\cite{BAYM1971225} yielded a polytropic constant \(K =(3.75785756 \times10^{-4} \pm 1.46\times10^{-5}) \, \text{MeV fm}^{-3(1-\gamma)}\) and an adiabatic index \(\gamma = 1.61131624  \pm 9.46 \times 10^{-3}\).The values were obtained with the {\it curvefit} function from SciPy package, employing the Levenberg-Marquardt algorithm.

 Despite the simplicity of this polytropic EOS, our results closely matched those obtained using the exact Douchin and Haensel (DH) for the inner crust \cite{sly4,Read:2008iy}. Consequently, for the approximations considered here, both inner crust EOS models produced indistinguishable results.

\begin{figure}[H]
    \centering
        \includegraphics[scale=0.3]{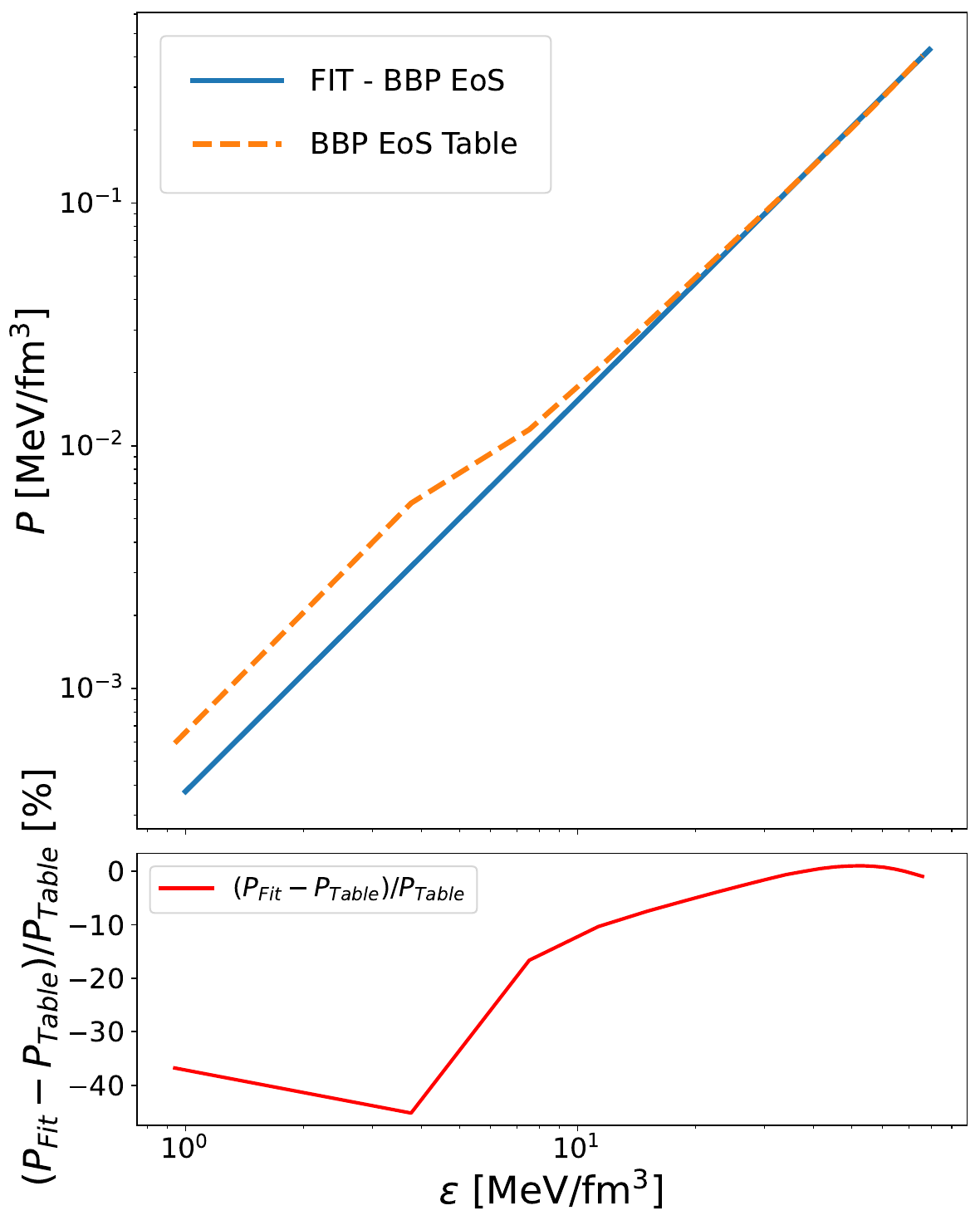}
        \caption{The top figure shows the comparison between the BBP EoS and its respective fit. The bottom figure shows the relative error between them. In the usual energy density crust-core transition for relativistic mean field models (from 62.5 MeV/fm³ to 86 MeV/fm³ \cite{Dutra2021}) our fit show less than $ |5|  \%$ relative error.}
        \label{fig:10}
    \end{figure}

\section{Modified Gravities, anisotropic pressures and dark matter}

As mentioned above, modified theories of gravity\cite{Berti:2015itd,HESS2020103809} can alter neutron star mass–radius relations by introducing extra terms (or by modifying) in the Tolman--Oppenheimer--Volkoff (TOV) equations, and often lead to an anisotropic pressure~\cite{Wojnar2016,daSilva2025,Rahmansyah2022,Mota:2019opp}. Additional sources of anisotropy in the r.h.s. arise in models involving scalar fields (e.g., boson stars), exotic solutions to Einstein's equations (such as wormholes or gravastars), configurations with charged matter~\cite{Horvat:2010xf}, magnetized stars~\cite{Paret_2015}, or exact solutions to the Einstein equations with an anisotropic component~\cite{Dev2002}.

\subsection{Neutron stars in the context of  $f(R,L_{m},T)= R +\alpha T L_{m}$ gravity model}

Evidently this vast subject is impossible to address properly within an Appendix, therefore we just provide some examples of the additional ingredients that may be involved in a detail modeling of compact stars.
As an example of modified gravity, the theory of $f(R,L_{m},T)= R +\alpha T L_{m}$\cite{Haghani:2021fpx} can be investigated in the context of NS following the Ref.\cite{Mota:2024kjb}, and using the analytical representation\footnote{Details are provided in Appendix C.} of the unified SINPA\cite{PhysRevD.106.023031} EoS. The modified TOV for $L_{m}=p$ is giving by\cite{Mota:2024kjb},

\begin{align}
\frac{dm}{dr} &= 4\pi r^2\epsilon + \frac{\alpha r^2}{2}\Bigl[\frac{\epsilon(5p - \epsilon)}{2} + p^2)\Bigr], \\[1.4em]
\frac{dp}{dr} &= -\frac{(\epsilon+p)\Bigl(4\pi r p + \frac{m}{r^2} + \frac{\alpha r}{4}(3p-\epsilon)p\Bigr)}
{\Bigl(1 - \frac{2m}{r}\Bigr)\Biggl[1 + \frac{\alpha p}{16\pi+\alpha(5p-\epsilon)} (1 - \frac{d\epsilon}{dp} ) \Biggr]}.
\end{align}
where $\alpha$ is a matter-geometry coupling constant (free parameter) and T= $- \epsilon + 3p$. In our calculations, $\alpha= x \mu_{1}$, where $\mu_{1}=1.46 \times 10^{10} ~ m^{2}$\cite{Mota:2024kjb}. As studied in Ref. \cite{Mota:2024kjb}, we will investigate the non-conservative effects ($\nabla^{\mu}T_{\mu\nu} \neq 0 $) related to  neutron star's surface radius and maximum mass. One must note that for $\alpha=0$, we retrieve the pure General Relativity or the standard TOV equations.

\begin{figure}[H]
    \centering
        \includegraphics[scale=0.4]{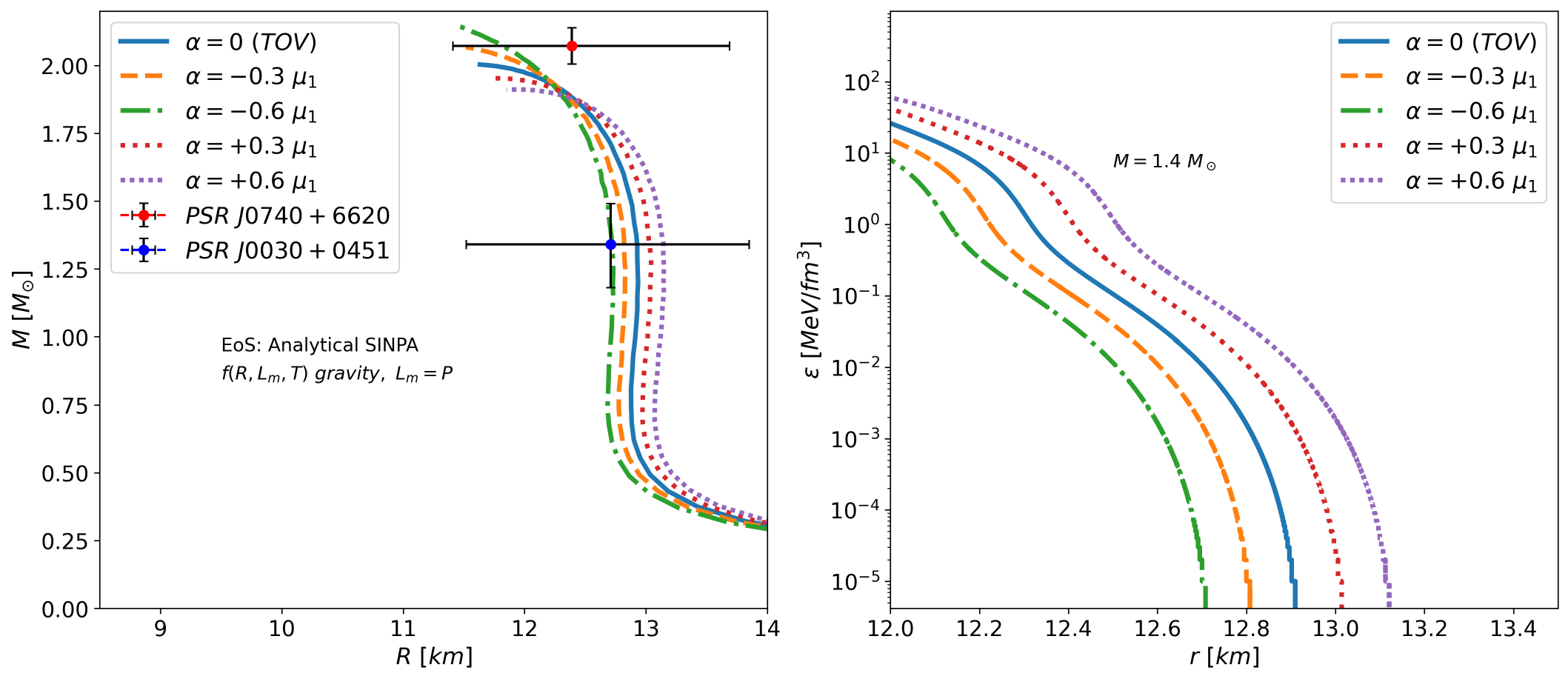}
        \caption{The left panel illustrates the dependence of the stellar sequence on $  \alpha  $. The right panel presents the density profile for a fixed mass of $1.4 ~ M_{\odot}$ corresponding to the same values of $  \alpha  $.}
        \label{fig:11}
    \end{figure}

One can see in the left panel of Fig. \ref{fig:11}, that for negative values of $\alpha$ the radius from stellar sequence decrease, and before achieving the maximum mass related to TOV solutions, goes above it. The opposite effect is seen for positive values of $\alpha$, i.e. increase in the stellar radius and reduction in the maximum mass related to TOV solutions. However, when considering $L_{m}=-\epsilon$, the modified TOV is not the same as the above and the results for the M-R relations changes drastically as shown in \cite{Mota:2024kjb} for the same EoS. There are similarities between the M–R relations obtained in the present case ($  L_m = -\epsilon  $) and those for the case with anisotropic pressure, as shown in Fig.~1 (bottom left) \cite{Biswas:2019gkw}.
In the right panel of Fig.~\ref{fig:11}, the surface radius varies depending on the sign (positive or negative) and the value of $  \alpha  $. 
Both results demonstrate the difficulties in breaking the degeneracies between the EoS and the free parameters associated with the modified gravity model.

A similar study\cite{Quartuccio:2024wnh} employed a specific functional form of $  f(R,T)  $ that was obtained by applying Gaussian process regression to measurements of the Hubble parameter. In the context of neutron stars, the authors adopted $  L_m = -p  $, while employing the following functional form:
$  f(T) = \alpha T^{2} + A \tanh[\lambda(T + T_{0})] + \beta T + \gamma  $.
To adequately describe a neutron star, three parameters from the original functional form must be modified, namely $  A  $, $  \beta  $, and $  \gamma  $, all of which should assume values close to zero. The authors employed a polytropic equation of state\footnote{$  K = 1.24 \times 10^{-4} \, \text{MeV} \, \text{fm}^{-3(1-\gamma)}  $ and $  \gamma = 2.0  $.} and found that, compared with GR solutions, the maximum mass of the neutron star (NS) is slightly increased.
The use of an analytical equation of state, together with a re-evaluation of the impact of the three parameters, can provide further insight into the functional form of $f(R,T)$. Furthermore, as noted in Ref.~\cite{Quartuccio:2024wnh}, most $  f(R,T)  $ functions proposed in the literature have been introduced on an {\it{ad hoc}} basis.

\subsection{Anisotropic pressures}

As shown in our results, the mass–radius relation shifts to the right (increased radius) with an increase in maximum mass, while the dimensionless tidal deformability decreases for the same positive value of the anisotropy parameter. This behavior aligns with that reported in several other  studies~\cite{Biswas:2019gkw,PhysRevD.106.103518,PhysRevD.85.124023} of neutron stars. However, for self-bound stars (strange quark stars), the tidal deformability increases for the same positive anisotropic parameter~\cite{PRETEL2024138375,Arbanil:2023yil,Lopes:2023phz}.

In this context, we exemplify additional uncertainties due to the impact of a quasi-local EoS~\cite{Horvat:2010xf} for the transverse pressure in the $\rcore$, total mass and  dimensionless tidal deformability ($\Lambda$) for the MPA1 EoS. In our calculations, only positive values of the anisotropy parameter $HB$ ($\alpha > 0$) were considered. Positive values increase the maximum mass, whereas negative values reduce it. 
Since the contribution from anisotropic pressures is expected to have only minor effects 
on the macroscopic structure of neutron stars, a small range for $\alpha = [0.1, 0.3]$ was chosen

\begin{figure}[H]
    \centering
    
    \begin{subfigure}{\textwidth}
        \centering
        \includegraphics[width=0.8\textwidth]{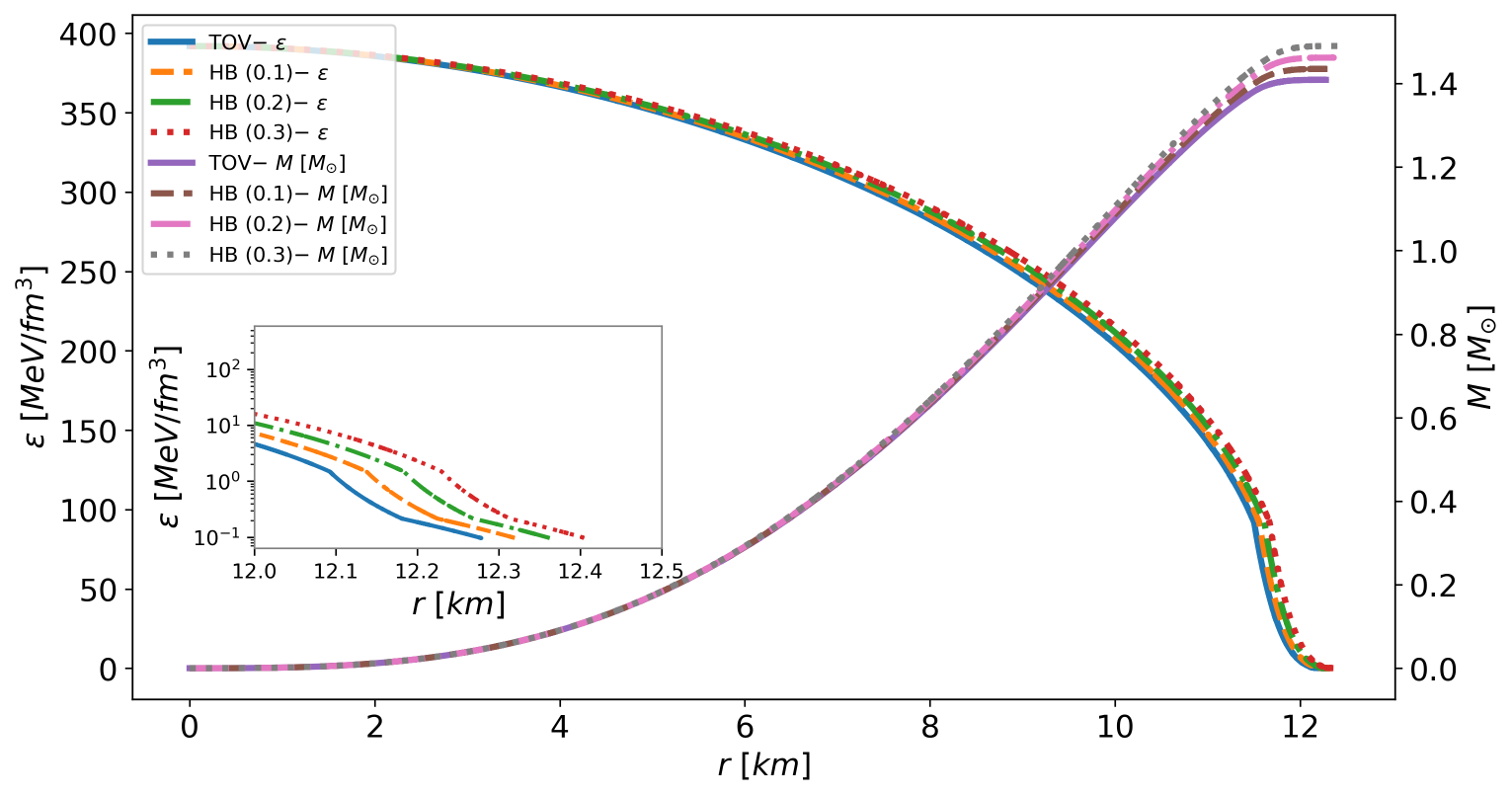}
        \caption{On the $y$-axis on the right (in $M_{\odot}$), we observe a small increase in the maximum mass when anisotropic pressure ($\alpha>0$) is included. On the $y$-axis on the left ($\epsilon$), the inclusion of this transverse component causes a small increase in $\rcore$. However, as shown in the inset plot, this small increase cannot be distinguished from isotropic solutions (TOV). }
        \label{fig:51}
    \end{subfigure}
    
    \vspace{1cm} 
    
    \begin{subfigure}{\textwidth}
        \centering
        \includegraphics[width=0.8\textwidth]{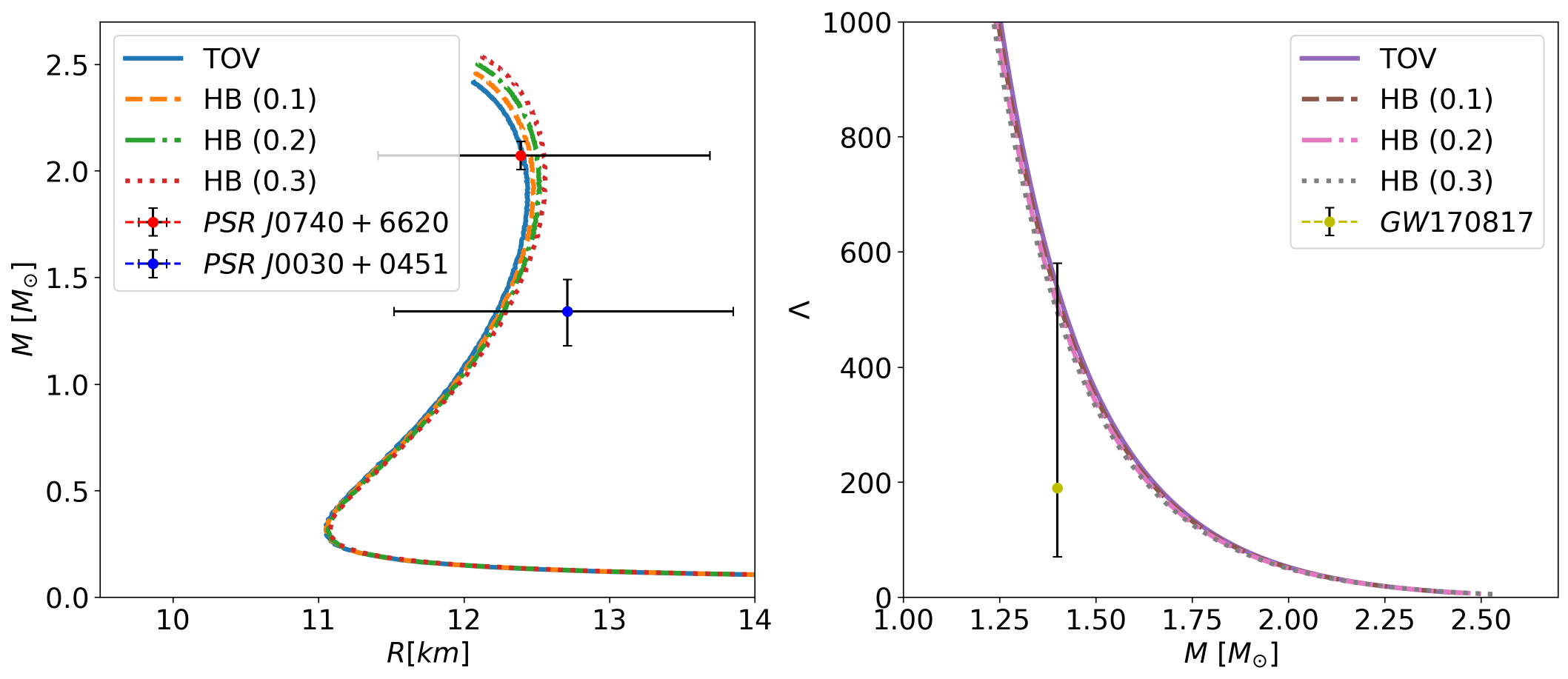}
    \caption{The left plot illustrates the effects of anisotropic pressure using the EoS HB in the mass-radius relations. Again, the effects of anisotropic pressure is only perceived in the maximum mass. In other words, the error bands from the observations of PSR~J0030+0451~\cite{riley2019nicer}  and PSR~J0740+6620~\cite{riley2021nicer} cannot disentangle the effects of isotropic and anisotropic pressures. The same holds for the right panel, but for the tidal deformability associated with gravitational event GW170817 \cite{LIGOScientific:2018cki}. }
        \label{fig:61}
    \end{subfigure}

    \caption{Effects of anisotropy pressures on mass-radius relations, dimensionless tidal deformability and $\rcore$ for the MPA1 EoS.}
  
    \label{fig:mpa1_ani}
\end{figure}

Another source of uncertainty is due to the equation of state with transverse pressure, commonly seen in the 
literature, i.e., Bowers and Liang\cite{Bowers1974}. The contributions to 
the mass--radius relations are more pronounced, as are those to the tidal 
deformability. However, for small values of the anisotropy parameter, both EoSs 
yield similar results. This can be seen in Fig.~3 \cite{PhysRevD.106.103518} where the authors used the parametrization IOPB-I based on relativistic mean field models. 

\subsection{Dark matter effects on neutron stars}

Similar effects to the case $\alpha < 0$ in the mass--radius relation can be achieved in
dark-matter-admixed neutron stars. Ref.~\cite{Zhou:2025dmy} considers a fermionic dark
matter model with particle mass $\mu$ and self-interactions. Using the two-fluid
formulation of the Tolman--Oppenheimer--Volkoff (TOV) equations, the authors show in
their Fig.~3 that, for $\mu = 2.0$~GeV and dark matter fractions $0 < f < 0.4$, both the
radius and the maximum mass decrease. The reduction of the core radius $r_\text{core}$
for a fixed neutron star mass of $1.4\,M_\odot$, $\mu=2.0$ ~ GeV and varying dark matter fractions
($0 < f < 0.3$) is displayed in their Fig.~5. As it stands, this type of contributions should 
be studied case by case, but it is clear that even a clean measurement of the radius with 
small ($\leq 100 \, m$ precision) error bars would still be plagued with them.

\section{Analytical representation for the Unified SINPA EoS} 

Analytical representations offer two significant advantages over tabulated data. First, they eliminate any ambiguity associated with interpolation, thereby enabling the exact and precise computation of derivatives. Second, they can be formulated in such a way that the thermodynamic relations are satisfied exactly\cite{Potekhin:2013qqa}. Following the analytical representation for the pressure\cite{Potekhin:2013qqa}, where $\xi = \log10(\rho/\mathrm{g\,cm^{-3}})$ and $\zeta = \log10(P/\mathrm{dyn\,cm^{-2}})$, the parametrization of $\log 10 (P(\rho))$ reads

\begin{equation}
\begin{aligned}
\zeta ={}&
\frac{a_1 + a_2\xi + a_3\xi^3}{1 + a_4\xi}
\bigl[\exp[a_5(\xi - a_6)] + 1\bigr]^{-1}
\\[1.2ex]
&+ (a_7 + a_8\xi)
\bigl[\exp[a_9(a_6 - \xi)] + 1\bigr]^{-1}
\\[1.2ex]
&+ (a_{10} + a_{11}\xi)
\bigl[\exp[a_{12}(a_{13} - \xi)] + 1\bigr]^{-1}
\\[1.2ex]
&+ (a_{14} + a_{15}\xi)
\bigl[\exp[a_{16}(a_{17} - \xi)] + 1\bigr]^{-1}
\\[1.2ex]
&+ \frac{a_{18}}{1 + [a_{19}(\xi - a_{20})]^2}
\\[1.0ex]
&+ \frac{a_{21}}{1 + [a_{22}(\xi - a_{23})]^2}.
\end{aligned}
\label{eq:3}
\end{equation}

The inverse of number density  in terms of mass density is giving by the following equation\cite{Potekhin:2013qqa},
\begin{equation}
\begin{aligned}
\frac{\tilde{\rho}}{n} ={}& 1 
+ (1 - f_2) \frac{c_1 \tilde{\rho}^{c_2} + c_3 \tilde{\rho}^{c_4}}{(1 + c_5 \tilde{\rho})^3} \\
&+ \frac{\tilde{\rho}}{c_6 + c_7 \tilde{\rho}^{c_8}} \, f_2,
\end{aligned}
\label{eq:7}
\end{equation}
where $f_2 \equiv \bigl[\exp(\xi - c_9) + 1\bigr]^{-1}$ and  $\tilde{\rho} =\frac{\rho}{1.66 \times 10^{15} \, \mathrm{g\, cm^{-3}}}$. The number density is in $(fm^{-3})$.

\begin{table}[H]
\centering
\caption{Pressure coefficients (left) and coefficients for the inverse number density relation (right)}
\label{tab:combined-coefficients}
\begin{minipage}{0.48\textwidth}
\centering
\small
\begin{tabular}{ccc}
\toprule
Coefficient & Value & $\sigma$ (standard deviation) \\
\midrule
$a_{1}$  & -279.7732515526258  & 11192.663348488148 \\
$a_{2}$  & -156.7548139559178  & 9058.553075061098  \\
$a_{3}$  & 11.640153157672952   & 552.8895662724724   \\
$a_{4}$  & 0.2756333987211074   & 13.234975481334612  \\
$a_{5}$  & -1.1685322469858783  & 0.628990529998443   \\
$a_{6}$  & 10.49004527894803    & 0.902535247503645   \\
$a_{7}$  & -95.04999623101027   & 4905.658658208611   \\
$a_{8}$  & -23.453455877490573  & 595.7205473562314   \\
$a_{9}$  & -0.9727097538430152  & 1.1959565571233983  \\
$a_{10}$ & 40.74495100762901    & 4903.340750564782   \\
$a_{11}$ & 25.19929261494951    & 595.2030151784851   \\
$a_{12}$ & -0.8582768308096772  & 0.42061341293929655 \\
$a_{13}$ & 11.452396304544925   & 3.2821474106239115  \\
$a_{14}$ & 526.5353075153471    & 280.27209534235783  \\
$a_{15}$ & -37.00413933368038   & 19.582615920210305  \\
$a_{16}$ & 2.919337481137843    & 0.09169302738496457 \\
$a_{17}$ & 14.392801240500598   & 0.029967119530583335\\
$a_{18}$ & 4.9131367821116925   & 24.238654408101063  \\
$a_{19}$ & 0.6139315820198409   & 0.33345776651646625 \\
$a_{20}$ & 10.50255271080746    & 0.05988265223339503 \\
$a_{21}$ & -26.295971340731658  & 14.262913098665067  \\
$a_{22}$ & 0.9559342177411017   & 0.041687960835618265\\
$a_{23}$ & 14.32531899426688    & 0.03633162648764279 \\
\bottomrule
\end{tabular}
\end{minipage}
\hfill
\begin{minipage}{0.4\textwidth}
\vspace{-5.4cm}
\centering
\small
\begin{tabular}{ccc}
\toprule
Coefficient & Value & $\sigma$ (standard deviation) \\
\midrule
$c_{1}$ & 25.75315147615395   & 1005.6273252554762  \\
$c_{2}$ & 2.6246144329415557  & 1.020170448735065   \\
$c_{3}$ & 5.915931643649342   & 102.86076716753215  \\
$c_{4}$ & 3.454931009431836   & 0.16389975105355703 \\
$c_{5}$ & 2.109560285007014   & 3.4806828731254646  \\
$c_{6}$ & 16.60292382028622   & 35.131457264242876  \\
$c_{7}$ & 0.3420501460192847  & 6.0951816810120585  \\
$c_{8}$ & 1.4846999138053802  & 4.890153625353264   \\
$c_{9}$ & 16.690791322095198  & 40.46915263961446   \\
\bottomrule
\end{tabular}
\end{minipage}
\end{table}

The results for the coefficients\footnote{The fitting method employed here is identical to that used for the polytropic EoS. A better optimization method can improve the results. } in Table \ref{tab:combined-coefficients} are,

\begin{figure}[H]
    \centering
        \includegraphics[width=0.9\textwidth]{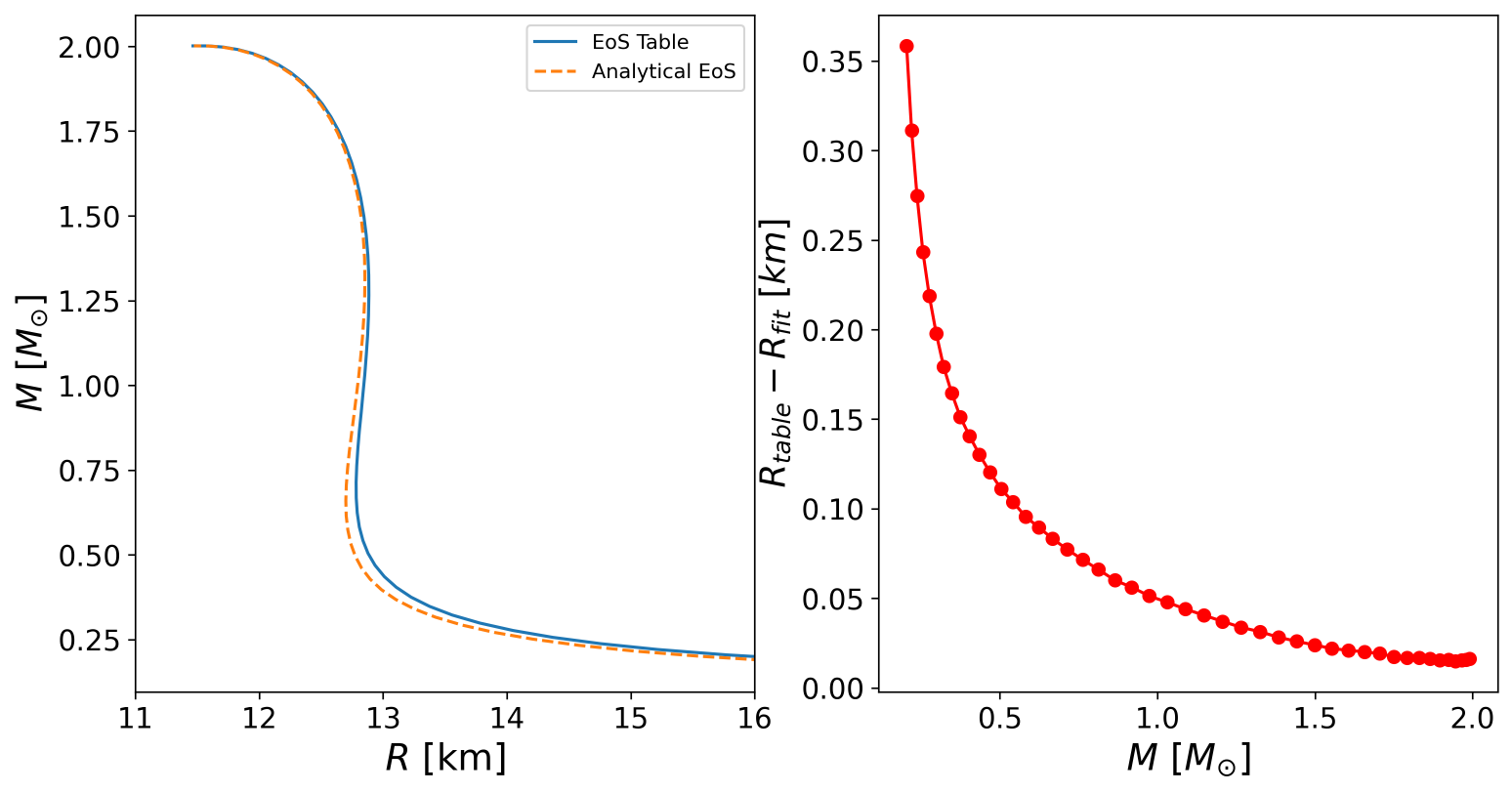}
        \caption{On the left plot, a small difference in radius is visible between the stellar sequences obtained using the Table EoS and the Analytical EoS. On the right plot, this radius difference is quantified as a function of neutron star mass. For neutron star masses greater than 1.0 $  M_{\odot}  $, the radius difference remains below 50 m.  }
        \label{danskjndaskj}
   \end{figure}

\nocite{*}
\bibliographystyle{unsrt}
\bibliography{sample} 

\end{document}